\documentclass[12pt]{article}

\usepackage{amsmath,amsthm,amsfonts,braket,algorithm,graphicx,multirow}
\usepackage{fullpage,authblk}

\theoremstyle{definition} 
\theoremstyle{definition} 
\newtheorem {theorem} {Theorem}

\newcommand{\kb}[1]{\mathbf{[#1]}}

\newcommand{\bk}[1]{\braket{#1|#1}}

\newcommand{\MR}{\texttt{Measure and Resend}}
\newcommand{\R}{\texttt{Reflect}}

\newcommand{\MTWO}{\texttt{MODE-2}}
\newcommand{\MTHREE}{\texttt{MODE-3}}

\newcommand{\prf}{p^{A\rightarrow B}}
\newcommand{\prr}{p^{A\rightarrow A}}
\newcommand{\pkey}{p^{\text{key}}}

\newcommand{\reff}{\widetilde{r}}

\floatname{algorithm}{Protocol}

\title{Semi-Quantum Key Distribution with High Quantum Noise Tolerance}
\author[1]{Omar Amer}
\author[1]{Walter O. Krawec\footnote{Email: \texttt{walter.krawec@uconn.edu}}}
\affil[1]{\small{Department of Computer Science and Engineering}\\\small{University of Connecticut}\\\small{Storrs, CT 06268 USA}}

\begin{document}
\maketitle
\begin{abstract}
Semi-quantum key distribution protocols are designed to allow two parties to establish a shared secret key, secure against an all-powerful adversary, even when one of the users is restricted to measuring and preparing quantum states in one single basis.  While interesting from a theoretical standpoint, these protocols have the disadvantage that a two-way quantum communication channel is necessary which generally limits their theoretical efficiency and noise tolerance.  In this paper, we construct a new semi-quantum key distribution (SQKD) protocol which actually takes advantage of this necessary two-way channel, and, after performing an information theoretic security analysis against collective attacks, we show it is able to tolerate a channel noise level higher than any prior SQKD protocol to-date.  We also compare the noise tolerance of our protocol to other two-way fully quantum protocols, along with BB84 with Classical Advantage Distillation (CAD).  We also comment on some practical issues involving semi-quantum key distribution (in particular, concerning the potential complexity in physical implementation of our protocol as compared with other standard QKD protocols).  Finally, we develop techniques that can be applied to the security analysis of other (S)QKD protocols reliant on a two-way quantum communication channel.
\end{abstract}


\section{INTRODUCTION}

Key distribution is the task of allowing two parties, typically referred to as Alice ($A$) and Bob ($B$), to agree on a shared secret key.  Such a key may be used in other cryptographic primitives such as encryption or authentication.  If $A$ and $B$ are restricted to communicating only through classical means, it is impossible to create a secure key distribution protocol, unless one makes computational assumptions on the power of the adversary Eve ($E$) (e.g., unless one assumes certain problems are computationally difficult to solve in a ``reasonable'' amount of time).  However, if we provide $A$ and $B$ with quantum communication capabilities, this impossibility result no longer holds.  Indeed, with \emph{Quantum Key Distribution} (QKD) protocols, it is possible for $A$ and $B$ to agree on a shared secret key which is secure against an all-powerful adversary.

Thus, if both parties $A$ and $B$ only have classical capabilities, information theoretic key distribution is impossible; if both parties have quantum capabilities, perfect security \emph{is} possible.  Is there a middle-ground?  In 2007, Boyer et al., \cite{SQKD-first} introduced the \emph{semi-quantum} model of cryptography in order to shed light on this question.  A protocol that is semi-quantum requires only one of the honest participants to be \emph{quantum capable} while the other participants are ``\emph{classical}.''  For the key-distribution problem, a semi-quantum key distribution (SQKD) protocol requires $A$ to be the quantum user (who is able to work with qubits in arbitrary manners) while the other user, $B$, is the classical user (who is restricted to working in a single basis or disconnecting from the quantum channel - we will discuss the exact capabilities of both users later in this paper).  We comment that this ``classical'' user as used in our protocol, though restricted from a theoretical point of view (in that he cannot perform certain measurements) may actually be more complicated to implement in practice due to his need to measure and resend quantum states (and, at an implementation level, still must be able to control and generate quantum states). Indeed, standard ``fully-quantum'' protocols are easier to implement with today's technology.

Our interest in studying this semi-quantum protocol, however, is from a theoretical perspective especially to see how security is affected with the classical user's theoretical limitation on not being able to perform certain measurements.  However, the techniques used here may be applicable to other, potentially practical, SQKD protocols such as the recently developed ``mirror'' protocol \cite{boyer2017experimentally}; we leave this practical study as interesting future work and restrict ourselves to a theoretical analysis in this paper.

\begin{figure}
  \centering
  \includegraphics[width=300pt]{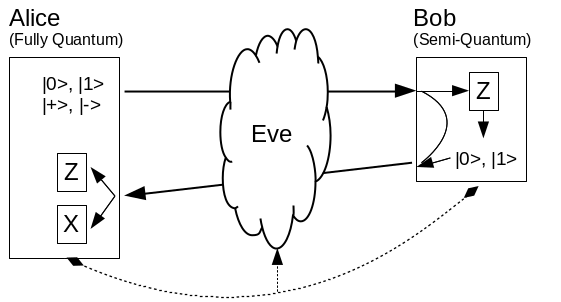}
  \caption{A standard SQKD protocol.  Alice begins by sending qubits, prepared in arbitrary bases, to Bob.  Bob, the ``classical'' or ``semi-quantum'' user is allowed only to work directly with the computational $Z$ basis, consisting of states $\ket{0}$ and $\ket{1}$, or he may disconnect from the channel and reflect everything back to $A$ (thus learning nothing about the qubit state).  When a qubit returns, $A$ is allowed to perform any quantum operation on the qubit.  Eve is allowed to attack both forward and reverse quantum channels; she may also intercept, but not tamper with, any message sent on the classical authenticated channel (depicted as a dashed line on the bottom of the diagram).}\label{fig:sqkd}
\end{figure}

Despite the theoretical interest to the semi-quantum model, there is one severe draw-back in that a two-way quantum communication channel is required for their operation (see Figure \ref{fig:sqkd}).  Such a channel allows a qubit to travel from the quantum user $A$, to the limited classical user $B$, then back to $A$.  This potentially introduces greater noise as a qubit must travel farther.  While favorable noise tolerances have been computed for some SQKD protocols now (see \cite{krawec2015security,krawec2016quantum,zhang2018security}; in particular, as shown in \cite{krawec2016quantum}, the original SQKD protocol of Boyer et al. \cite{SQKD-first} has a noise tolerance which exactly matches that of BB84 \cite{QKD-BB84} operating over two independent channels), this limitation has been a bottleneck in these results.

In this work we turn this disadvantage into an advantage by proposing a new protocol, based on a modified version of the original SQKD protocol constructed by Boyer et al., \cite{SQKD-first}.  This new protocol takes advantage of information transmitting on \emph{both} channels to improve noise tolerance.  In fact, we will show that the noise tolerance for our new protocol can be as high as $26\%$ over certain channels (for other channels, the maximum ranges between $12.5\%$ and $17.8\%$).  This is higher than any other semi-quantum protocol which has thus far been analyzed.  This is also higher than, to our knowledge, other two-way fully-quantum (i.e., not semi-quantum, but where both parties are quantum \emph{and} rely on a two-way quantum channel) protocols (such as LM05 \cite{QKD-LM05} and the PingPong Protocol \cite{QKD-PingPong} which, as shown in \cite{QKD-TwoWaySecure} achieve at most $12\%$).  It is not as high as BB84 with Classical Advantage Distillation (which can have a theoretical noise tolerance of $27.6\%$ \cite{CAD,BB84-CAD,chau2002practical}), but it does compare favorably as we discuss later in our evaluations.

We make several contributions in this work.  First, in Section II we propose a new protocol that successfully takes advantage of the two-way nature of the quantum communication channel in order to improve noise tolerance.  This is the first time such a task has been achieved in the semi-quantum model and the techniques used may be applicable to other (S)QKD protocols.  Second, in Section III, we perform an information theoretic security analysis of the protocol computing its key-rate and noise tolerance against collective attacks.  To do so, we also extend the technique of mismatched measurements \cite{QKD-Tom-1,QKD-Tom-2,krawec2016quantum} to support additional information from a third basis on a two-way channel; these techniques can be applied to other SQKD (or even other fully-quantum protocols using a two-way channel) extending the applicability of the results in this paper; this is achieved in Section III.B.  We compare our results to standard QKD protocols in Section IV. Finally, in Section V, we also consider channel loss.

Our new protocol is, in a way, a semi-quantum version of BB84 with Classical Advantage Distillation (CAD) \cite{CAD,BB84-CAD,chau2002practical}.  Thus, we are particularly interested in comparing our protocol to that of BB84 with CAD.  We show that our protocol's noise tolerance \emph{and efficiency} is higher for certain CAD techniques and, while investigating this, we discover a very interesting property of our protocol, namely that $E$'s uncertainty is higher in our two-way protocol than it would be in BB84 with CAD for certain, realistic, channel scenarios.  Furthermore, our protocol has the advantage that two-way classical communication over the authenticated channel (an expensive resource), necessary for CAD, is not required in our protocol.  Finally, while to utilize CAD with BB84, both parties must do additional computations, in our protocol only the fully quantum user must take additional actions (though we note again that in practice it may be that with current technology, the classical user’s actions may be more difficult to realize still than the actions of Bob in BB84).  These results and observations may spur future research in the design of advantage distillation protocols operating over \emph{quantum} channels as opposed to \emph{classical} ones (such as standard CAD techniques).

\subsection{Semi-Quantum Cryptography}

Since its creation in 2007 by Boyer et al., in \cite{SQKD-first,SQKD-second}, SQKD research has expanded greatly to include newer protocols, new primitives beyond key-distribution, and new proof techniques to argue about security.

In the semi-quantum model a two-way quantum channel connects two users $A$ and $B$ (see Figure \ref{fig:sqkd}).  One participant, typically $B$, is forced to be ``classical'' or ``semi-quantum'' in nature in that he may only directly operate in the computational $Z$ basis (consisting of states $\ket{0}$ and $\ket{1}$); or he may choose to ``disconnect'' from the quantum channel.  The other user, $A$, may be ``fully quantum'' and thus prepare, measure, and interact with quantum resources in arbitrary ways.

More specifically, these protocols typically begin with $A$ producing quantum bits and sending them to the limited, ``classical'', user $B$.  This user, on receiving a qubit, is allowed to perform one of two operations:
\begin{enumerate}
  \item $\MR$: If $B$ chooses this option, he will perform a $Z$ basis measurement on the qubit resulting in outcome $\ket{r}$ for $r \in \{0,1\}$.  He will then send a qubit to $A$ in the state $\ket{r}$.  $B$ is only allowed to measure and send in the $Z$ basis.
  \item $\R$: In this case, $B$ simply ignores the incoming qubit and reflects it back to $A$.  The qubit remains undisturbed, however $B$ learns nothing about its state.
\end{enumerate}

One trend among SQKD research is to produce new protocols requiring even fewer quantum capabilities of the two users.  Originally SQKD protocols placed resource restrictions only on the ``classical user'' Bob.  However, in 2009 Zou et al., \cite{SQKD-lessthan4} introduced several new SQKD protocols utilizing resource restrictions also on the quantum user in that she could not prepare arbitrary states; indeed one such protocol permitted $A$ to only prepare a single, publicly known state (called a \emph{single-state} protocol).  Other works have also analyzed the case when the quantum user is restricted to sending only one (or few) states \cite{SQKD-Single-Security,zhang2018security}.  Limitations on $A$'s measurement capabilities, in addition to her source preparation abilities, are also possible \cite{krawec2017limited}.  Researchers have also considered further restricting the capabilities of the classical user $B$ \cite{zou2015semiquantum}.

Beyond standard point-to-point key distribution, the notion of \emph{mediated semi-quantum key distribution} was introduced in \cite{SQKD-MultiUser}, with an improved protocol introduced in \cite{liu2018mediated}.  In this model, a fully-quantum server (who prepared and measured quantum states) was utilized allowing two ``classical'' users to establish a shared secret key with one another.  Rigorous security proofs (including noise tolerance computations) exist for some protocols, even if the server is adversarial \cite{SQKD-MultiUser}.

As far as security is concerned, most SQKD protocols are proven to be \emph{robust}, a notion introduced in \cite{SQKD-first}.  A protocol is robust if, for any attack causing an adversary to learn non-zero information, there is a non-zero probability that the adversary may be detected.  More rigorous notions of security have been proven for some SQKD protocols.  In \cite{krawec2016quantum}, the noise tolerance and key-rate of the original SQKD protocol has been shown to match that of BB84 assuming the technique of \emph{mismatched measurements} is used. Mismatched measurements, which utilize statistics such as the probability of measuring a $\ket{0}$ if a $\ket{+}$ was initially sent, are a useful technique used to improve the noise tolerances of many one-way QKD protocols \cite{QKD-Tom-1,QKD-Tom-2}; in \cite{krawec2016quantum} the technique was extended to two-way semi-quantum protocols for the first time.  If mismatched measurements are not used, the current best noise tolerance for the original SQKD protocol was found to be $6.14\%$ \cite{krawec2018key} (though the proof technique used in that paper, involving a reduction to an entanglement based one-way protocol, only found a lower-bound; it is still an open question as to whether the full $11\%$ noise tolerance can be found without mismatched measurements).  In \cite{zhang2018security} the noise tolerance of a single-state protocol originally introduced in \cite{SQKD-lessthan4} (and proven only to be robust in that original paper) was found to be $9.65\%$.

Beyond key distribution itself, the semi-quantum model has been used for other cryptographic tasks including secret sharing \cite{SQKD-secret1,SQKD-secret2,SQKD-secret3,SQKD-secret-efficient}, direct communication \cite{zou2014three,shukla2017semi,yan2018semi}, and quantum private comparison \cite{SQKD-comp1,SQKD-comp2,SQKD-comp3,SQKD-comp4}.  There has also been work recently in analyzing SQKD protocols in more practical settings (where it is impossible for $B$ to accurately perform the $\MR$ operation as he cannot prepare a photon in exactly the same state it was received) \cite{boyer2017experimentally,krawec2018practical,SQKD-prac3}.

However, of all the semi-quantum protocols in existence, none, to our knowledge, actually \emph{take advantage} of the two-way quantum channel (beyond, of course, permitting $B$ to be classical in nature).  In this work, we introduce a new protocol, modified from the original Boyer et al., protocol \cite{SQKD-first} and conduct an information theoretic security analysis computing its key-rate and noise tolerance.  We show that our modified protocol can actually strategically use the two-way quantum channel to increase noise tolerance (at a potential cost of decreased efficiency).  In some way, our protocol uses the two-way \emph{quantum} channel to run a simulated \emph{classical} advantage distillation (CAD) protocol \cite{CAD,BB84-CAD}.  However, when we later compare our results with CAD applied to BB84, we discover some rather surprising results.

\subsection{Practicality of Semi-Quantum Key Distribution}

As mentioned in the introduction, our primary interest in SQKD is theoretical, namely to help study the question ``how quantum must a protocol be to gain an advantage over its classical counterpart?''  From a practical stand-point, there are numerous technical challenges that seem to complicate implementation.  At its base, classical-Bob must be able to choose $\R$ or $\MR$, and it is the latter operation which creates some difficulties.  Indeed, if we consider a photonic implementation, $B$ must, on choosing this operation, perform a destructive measurement using a photon detector.  Then, subsequently, he must re-prepare a fresh qubit (in this case, a new photon).  Such operations lead to potential attacks such as the photon tagging attack \cite{SQKD-photon-tag} (see, also, the subsequent comment in \cite{SQKD-photon-tag-comment} for a potential countermeasure).  Furthermore, this operation also requires $B$ to be able to ensure that a single photon enters his device as, otherwise, $E$ could ``spam'' $B$'s device with $2$ or more photons and count the number leaving - if this number differs, $E$ can guess that $B$ choose $\MR$; otherwise he must have chosen $\R$.  While $B$ could police such activity with, perhaps, cascading beamsplitters and photon detectors, it creates another implementation challenge.

We do not address these implementation challenges in this work and the protocol we create, though secure in our theoretical analysis, would be susceptible to these attacks and would therefore require greater complexity in implementation to attempt to secure it against these.  Indeed, standard BB84 would probably be easier to implement in practice than our protocol.  Instead the advantage to our system is in its theoretical analysis.

However, that is not to say that semi-quantum is not practical to implement.  While, perhaps with current-day technology, standard one-way protocols are still easier to engineer, there is still a potential practical benefit to studying these limited protocols.  The issues of $\MR$ discussed above (namely, photon tagging and the multi-photon-counting attack as described) only apply to those protocols implementing the theoretical version of the $\MR$ operation (consisting of a measurement and recreation of the qubit).  Some newer SQKD protocols, in particular the ``mirror'' protocol introduced in \cite{boyer2017experimentally}, seek to mitigate these attacks by not requiring $B$ to ever destructively measure and then prepare a fresh photon (while still maintaining a mathematical equivalence to the semi-quantum model).  The techniques used in \cite{boyer2017experimentally} may be applicable to other semi-quantum protocols, including, perhaps, the one we describe in this work, making semi-quantum a practical possibility.  We feel that, even though with today's technology, standard ``fully-quantum'' protocols are easier to implement, there is still important practical research to studying semi-quantum protocols beyond their pure theoretical interest.

Finally, beyond approaches used by the mirror protocol (which is discrete variable), there may be possibilities in designing continuous-variable (CV) semi-quantum protocols.  Though it is unclear what, exactly, a ``classical'' user would be capable of in a CV model (the exact theoretical model would have to be developed), recent work in standard (i.e., not semi-quantum) CV protocols operating over two-way quantum channels  have shown several practical and theoretical benefits \cite{CV2,CV4,CV5,CV6}. Increased secure communication rates are possible \cite{CV-floodlight} and also increased noise tolerance \cite{CV1}.  In \cite{CV3}, it was shown that two-way CV protocols can be more secure than one-way protocols when dealing with preparation noise.  Such techniques could lead to efficient semi-quantum systems and may be an alternative approach towards creating practical systems of this nature.  However, these issues we leave as interesting future work and are outside the scope of this paper.

\subsection{Notation}

Given a random variable $A$, we define $H(A)$ to be the Shannon entropy of $A$.  Namely, if $A$ takes value $i$ with probability $p_i$, for $i = 1, \cdots n$, then:
\[
H(A) = H(p_1,\cdots,p_n) = \sum_ip_i\log p_i,
\]
where all logarithms in this paper are base two.  If $A$ takes only two values, we write $H(p_1)$ to mean $H(p_1,1-p_1)$.  We denote by $H(A|B)$ to be the conditional Shannon entropy defined $H(A|B) = H(AB) - H(B)$, where $H(AB)$ is the joint entropy of $AB$ defined in the obvious way.

We denote by the \emph{computational basis}, or $Z$ basis, states of the form $\ket{0}$ and $\ket{1}$.  The $X$ basis consists of states of the form $\ket{\pm} = \frac{1}{\sqrt{2}}(\ket{0} \pm \ket{1})$, and finally the $Y$ basis consists of $\ket{j_Y} = \frac{1}{\sqrt{2}}(\ket{0} + i(-1)^j\ket{1})$.

Given density operator $\rho_{AB}$ acting on Hilbert space $\mathcal{H}_A\otimes\mathcal{H}_B$, we write $\rho_B$ to mean the operator resulting from the partial trace over the $A$ system.  Also, given $\ket{\psi} \in \mathcal{H}_A$, we write $\kb{\psi}_A$ to mean $\ket{\psi}\bra{\psi}_A$.  If the context is clear, we forgo writing the subscript.  A state $\rho_{AB}$ is said to be a \emph{classical-quantum} state (or cq-state) if it can be written in the form: $\rho_{AB} = \sum_ap_a\kb{a}\otimes\rho_B^{(a)}$ for some orthonormal basis $\{\ket{a}\}$.

Given density operator $\rho_{A}$ we define $S(A)_\rho$ to mean the von Neumann entropy of $\rho_A$ (i.e., $S(A)_\rho = S(\rho_A) = -tr(\rho_A\log\rho_A)$).  We write $S(A|B)_\rho$ to mean the conditional von Neumann entropy of the $A$ register of state $\rho_{AB}$ conditioned on the $B$ register, namely: $S(A|B)_\rho = S(AB)_\rho - S(B)_\rho$.  If the context is clear, we will forgo writing the subscript ``$\rho$.''


Finally, we will use the following theorem from \cite{krawec2016quantum}:
\begin{theorem}\label{thm:entropy} (From \cite{krawec2016quantum}):
Let $\rho_{AE}$ be a cq-state of the form:
\[
\rho_{AE} = \frac{1}{N}\kb{0}_A\otimes\left(\sum_{i=1}^{N_0}\kb{E_i^0}\right) + \frac{1}{N}\kb{1}_A\otimes\left(\sum_{i=1}^{N_1}\kb{E_i^1}\right),
\]
where the $\ket{E_i^j}$ are arbitrary (not necessarily normalized nor orthogonal) vectors in $\mathcal{H}_E$, then:
\begin{equation}
S(A|E)_\rho \ge \sum_{i=1}^{\min(N_0,N_1)}\left(\frac{\bk{E_i^0} + \bk{E_i^1}}{N}\right)\cdot\left(H\left[\frac{\bk{E_i^0}}{\bk{E_i^0} + \bk{E_i^1}}\right] - H[\lambda_i]\right),
\end{equation}
where:
\begin{equation}
\lambda_i = \frac{1}{2}\left(1 + \frac{\sqrt{(\bk{E_i^0} - \bk{E_i^1})^2 + 4Re^2\braket{E_i^0|E_i^1}}}{\bk{E_i^0} + \bk{E_i^1}}\right).
\end{equation}
\end{theorem}

\subsection{General QKD Security}

A QKD protocol typically operates in two stages: a \emph{quantum communication stage} and a classical \emph{post-processing stage}.  The first utilizes the quantum channel, along with the authenticated classical channel, in order for $A$ and $B$ to output a \emph{raw-key} which is partially correlated and partially secret.  The second stage, using the authenticated channel, consists of an error correction protocol (leaking additional information to Eve) and privacy amplification.  For more information on these two standard processes, the reader is referred to the survey \cite{QKD-survey}.

Before error correction and privacy amplification, however, there are other classical protocols that may be run at this stage.  One interesting approach is to run a \emph{classical advantage distillation} (CAD) \cite{CAD} protocol which serves to increase $A$ and $B$'s raw key correlation.  CAD has been applied to several protocols, including BB84, and has been shown to increase noise tolerance significantly.  We will revisit CAD applied to BB84 later in this work when evaluating our new protocol and comparing with current state-of-the-art protocols.

One important computation in any QKD protocol security proof is its \emph{key-rate}.  If $N$ is the size of the raw-key after the quantum communication stage, and $\ell(N) \le N$ is the size of the secret key after privacy amplification, then the key-rate is defined to be:
\[
r = \frac{\ell(N)}{N}.
\]
Another important variant of this ratio is the \emph{effective} key-rate which takes into account that not all iterations of the quantum communication stage yield a valid raw key bit. We extend the definition here to also take into account the total number of qubits prepared by the protocol (normally, for one-way protocols such as BB84, there is only one qubit per iteration so this extension is meaningless; however with a SQKD protocol there are actually two qubits per iteration prepared).  This effective key rate better measures the efficiency of a QKD protocol and is defined as:
\[
\reff = \frac{\ell(N)}{K},
\]
where $K$ is the number of qubits sent.  For an SQKD protocol, it is not difficult to see that $N$, the expected raw-key size, is simply:
\[
N = K\cdot p_{acc} \cdot \frac{1}{2},
\]
where $p_{acc}$ is the probability that, on any particular iteration of the quantum communication stage, that iteration yields a valid raw-key bit. Thus:
\begin{equation}\label{eq:eff-keyrate}
\reff_{SQKD} = \frac{1}{2}\cdot p_{acc}\cdot r.
\end{equation}


In this work we consider collective attacks where Eve attacks each iteration in an i.i.d. manner, but is allowed to postpone her measurement to any future point in time \cite{QKD-survey}.  With very few exceptions for certain ``nice'' protocols, most QKD security proofs are performed assuming collective attacks.  Usually, at least in the asymptotic scenario, security against collective attacks implies security against general attacks for protocols which are permutation invariant \cite{QKD-symmetric,QKD-general-attack,QKD-general-attack2}.

Under collective attacks, it was shown in \cite{QKD-Winter-Keyrate,QKD-renner-keyrate} that the following equation, known as the \emph{Devetak-Winter Key Rate}, holds:
\begin{equation}\label{eq:dw-keyrate}
r = \lim_{N\rightarrow \infty}\frac{\ell(N)}{N} = \inf[S(A|E) - H(A|B)],
\end{equation}
where the infimum is over all collective attacks which induce the observed channel statistics (e.g., noise).  The entropy computations are done over cq-states describing the protocol's raw-key output bit and Eve's quantum memory for one iteration.  Computing $H(A|B)$  is generally trivial, instead it is the computation of the von Neumann entropy $S(A|E)$ that requires great effort in a QKD security proof.

Once a key-rate expression is derived, one generally wishes to evaluate it.  Naturally, were the protocol run in practice, the noise values would be observed directly.  Here, however, we wish to evaluate our key-rate under certain ``reasonable'' noise scenarios.  Perhaps the most common noise scenario considered is a symmetric attack modeled by a depolarization channel:
\begin{equation}\label{eq:sym-noise}
\mathcal{E}_Q(\rho) = (1-2Q)\rho + QI.
\end{equation}
In particular, the observable noise is $Q$ while any mismatched measurement events (such as a $\ket{+}$ being measured as a $\ket{0}$ after passing through this channel) are $1/2$.  Since such statistics are observable for the protocols considered in this work, this may even be enforced.

\section{OUR PROTOCOL}\label{section:new-prot}
Our protocol is a semi-quantum one and, as with all (S)QKD protocols, there are two primary stages: the quantum communication stage, and the classical post-processing stage.  The first uses the two-way quantum channel and the authenticated classical channel to establish a raw-key while the second estimates the noise in the channel and, assuming this is not ``too high'' (to be discussed) then processes this raw-key to output a shared secret key.  Our protocol may be run in one of two modes depending on the quantum user $A$'s capabilities: in $\MTWO$, only two bases are used (the $Z$ and the $X$ basis), similar to four-state BB84, whereas in $\MTHREE$, three bases are used by $A$ ($Z$, $X$, and $Y$), similar to the six-state BB84 - of course, regardless of $A$'s capabilities, $B$ can only measure and send in the $Z$ basis;  note there is no advantage to defining a higher-basis mode.  We will analyze the security of our protocol in both modes.  The exact protocol is described in detail below:

$ $\newline
\textbf{Public Constants: }
\begin{itemize}
  \item \texttt{Mode} = $\MTWO$ or $\MTHREE$, specifying the mode of operation (based on $A$'s quantum capabilities, namely whether she can work with $2$ or $3$ bases).
  \item $p,q \in (0,1)$, probability values for certain choices.
\end{itemize}
$ $\newline
\textbf{Quantum Communication Stage:}
This stage repeats the following process until a sufficiently large raw-key is produced.
\begin{enumerate}
  \item \textbf{Alice's Preparation:}
  \begin{itemize}
    \item With probability $p$, $A$ sets an internal private register $b_A = Z$; otherwise, with probability $1-p$, she sets $b_A$ to be $X$ if $\texttt{Mode} = \MTWO$ or she sets $b_A$ to be $X$ or $Y$ (with probability $(1-p)/2$ each) if $\texttt{Mode} = \MTHREE$.
    \item If $b_A = Z$, $A$ sets an internal private register $k_A$ to be $0$ or $1$ with uniform probability and she sends the computational qubit state $\ket{k_A}$ to $B$.
    \item Otherwise, if $b_A = X$, $A$ sends the state $\ket{+}$ to $B$; or, if $b_A = Y$ (which only happens when $\texttt{Mode} = \MTHREE$), she sends the state $\ket{0_Y}$ to $B$.  \emph{Note that $A$ does not need to prepare and send states of the form $\ket{-}$ or $\ket{1_Y}$.}
  \end{itemize}

  \item \textbf{Bob's Operation:}
  \begin{itemize}
    \item When $B$ receives a qubit, he will choose to either $\MR$ or to $\R$, saving his choice in a private internal register $\texttt{Choice}_B$.  If he chooses $\MR$, he saves his measurement result (a $0$ or a $1$ as he can only measure in the $Z$ basis) in a private register $k_B$.  The probability of choosing $\MR$ is $p$ (though independent of $A$'s choice).
  \end{itemize}

  \item \textbf{Alice's Measurement:}
  \begin{itemize}
    \item $A$ will choose a basis to measure in, randomly, following the same distribution as in step 1 (\emph{however, the basis choice here will be independent of her initial choice thus allowing for mismatched measurements}).  Let $b_A'$ be the register storing her basis choice.  $A$ will then perform a measurement in this basis saving the result in a register $m_A$ (which may take on any value in the set $\{0,1,+,-,0_Y,1_Y\}$ based on the measurement outcome).
    \item If $b_A = b_A' = Z$ and if $m_A = k_A$ (i.e., she measures the same state she sent in the $Z$ basis), $A$ sets an internal register $\texttt{Accept} = 1$; otherwise it is set to $0$.
    \item With probability $1-q$, $A$ will set an internal register $\texttt{Test} = 1$; otherwise it is set to $0$.
  \end{itemize}

  \item \textbf{Communication:} \emph{(Note this step may be performed ``in bulk'' for all iterations, after performing the above steps for sufficiently long.)}
  \begin{itemize}
    \item Using the authenticated channel, $A$ discloses $(b_A, \texttt{Accept}, \texttt{Test})$ and $B$ discloses $\texttt{Choice}_B$.
    \item If $\texttt{Accept} = 1$, $\texttt{Test} = 0$, and $\texttt{Choice}_B = \MR$, then they will use this iteration's results to contribute towards their raw key.  Such an iteration is called a \emph{key-distillation} iteration as it successfully adds to the raw-key length.  In particular $A$ uses $k_A$ as a new raw-key bit while $B$ uses $k_B$.
    \item Otherwise, if the above condition is not true, this iteration is not used and instead it will be used later to determine statistics on the quantum channel noise (in particular, $B$ will send $k_B$ in this case).  Note that mismatched results (when $A$ chooses different preparation and measurement bases) are \emph{not} discarded.
  \end{itemize}
\end{enumerate}

Following the quantum communication stage, and assuming the noise level is low enough (which we will compute), error correction followed by privacy amplification will output a secret key.  These are standard processes and for more information, the reader is referred to \cite{QKD-survey}.

We note that the \textbf{Communication} stage, listed above, can actually be completed ``in bulk'' after the quantum communication stage is completed.  Also, in the asymptotic setting, the choice of $p$ and $q$ may be made arbitrarily close to $1$ thus increasing the protocol efficiency (as was done, for example, with BB84 in \cite{QKD-BB84-Modification} and also semi-quantum protocols in \cite{krawec2015security}).  Note that in the finite key setting, the choice of $p$ and $q$ would be very important (as they will lead to differing efficiency and differing sample sizes for error estimation).  However, in this work, we consider only the asymptotic scenario.

\section{SECURITY ANALYSIS}\label{section-security}

Our goal is to compute a bound on the Devetak-Winter key-rate expression (Equation \ref{eq:dw-keyrate}) as a function only of observable statistics.  That is, we compute a bound as a function only on parameters that may be observed from iterations which are \emph{not} used for key-distillation in our protocol (see the \textbf{Communication} stage of our protocol).  First, however, we require a description of the density operator, describing $A$, $B$, and $E$'s systems, for all key-distillation iterations (as those are the iterations from which error correction and privacy amplification are run, thereby leading to a secret key).

Our security analysis is organized as follows: First we will derive the necessary density operator required to analyze our protocol and compute the key-rate.  Using this, we will derive a lower-bound on the conditional von Neumann entropy $S(A|E)$ as a function of several inner-products resulting from $E$'s attack operator.  In subsection A we will show how to bound these inner-products in the case $\MTWO$ is used - this will be based on prior work in \cite{krawec2016quantum}.  In subsection B we will derive new methods to attain tighter bounds on these inner-products in the case $\MTHREE$ is used.  The methods derived in that section are general purpose and may be applied to other (S)QKD protocols.  This section will end with a general method to optimize $S(A|E)$ for each mode; we use this in Section IV to actually evaluate our bounds.

We analyze collective attacks in this paper so as to immediately compare our results with those in \cite{BB84-CAD} where Classical Advantage Distillation (CAD) \cite{CAD} was applied to BB84.  That work computed mutual quantum information as functions of collective attacks and so the most natural comparison would be to compute similar expressions when $E$ uses collective attacks.  While, in \cite{BB84-CAD}, the results of course extend to general attacks (through standard techniques) and, normally, collective attacks imply security against general attacks for permutation invariant protocols (it is not difficult to make our protocol permutation invariant in the usual way by having $A$ and $B$ permute their raw-key bits using a randomly chosen permutation \cite{QKD-renner-keyrate}) \cite{QKD-general-attack,QKD-general-attack2}, a rigorous proof of security for general attacks is outside the scope of this work.  Nonetheless, we suspect that the results and computations, besides being of interest to compare with the BB84+CAD results, will extend to the general security setting in the usual way.  Furthermore, our results here can also be used for finite-key analyses (where collective attacks are often studied) and that may also be interesting to study as future work.

Note in this section, we are considering a loss-less channel; in a later section we will extend this analysis to deal with loss.  In this case, a collective attack against a semi-quantum protocol, may be modeled without loss of generality as a pair of unitary attack operators $(U_F, U_R)$ each acting on $\mathcal{H}_T\otimes\mathcal{H}_E$, where $\mathcal{H}_T$ is the two-dimensional Hilbert space modeling the qubit in transit between $A$ and $B$ (and later between $B$ and $A$) while $\mathcal{H}_E$ is the Hilbert space modeling $E$'s quantum memory (whose dimension is arbitrary).  On each iteration, after $A$ sends a qubit initially to $B$ in the forward channel, $E$ will apply $U_F$; when the qubit returns from $B$ to $A$, $E$ will apply $U_R$ (acting on the same space that $U_F$ acted on, thus the action of $U_R$ may depend on $U_F$).  There are no assumptions on measurement strategy in a collective attack; indeed $E$ is free to postpone her measurement to any future point in time and is allowed to make, later, any arbitrary optimal \emph{coherent} measurement of her entire memory through the protocol's operation.

At the start of each iteration, we may assume $E$ prepares a fresh ancilla in some pure state known to her: $\ket{\chi}_E$.  From this, we may describe $U_F$ and $U_R$'s action as follows:
\begin{align}
U_F\ket{0,\chi}_{TE} &= \ket{0,e_0} + \ket{1,e_1}\\
U_F\ket{1,\chi}_{TE} &= \ket{0,e_2} + \ket{1,e_3}\notag\\
U_R\ket{i,e_j}_{TE} &= \ket{0,e_{i,j}^0} + \ket{1,e_{i,j}^1}\notag.
\end{align}

Our goal, now, is to construct a density operator $\rho_{ABE}$ describing one iteration of the protocol in which a key bit was successfully distilled.  The entire run of the protocol, for all key-distillation iterations, will then be $\rho_{ABE}^{\otimes N}$ where $N$ is the size of the raw-key.  Once this state is computed, we will then compute (or rather bound) $S(A|E)_\rho$.

Conditioning on an iteration being used for key-distillation, $A$ must send a $\ket{0}$ or $\ket{1}$, saving the value in an internal register $k_A$.  $E$ then attacks the traveling qubit using $U_F$ and sends the qubit in $\mathcal{H}_T$ to $B$ who performs a $Z$ basis measurement (since he must choose $\MR$ for this iteration to potentially yield a new raw-key bit).  His measurement result is saved in his private register (which, below, we denote as simply $B$).  The resulting density operator, thus far, is easily found to be:
\[
\frac{1}{2}\kb{0}_{k_A}\otimes(\kb{0}_B\otimes\kb{e_0}_E + \kb{1}_B\otimes\kb{e_1}) + \frac{1}{2}\kb{1}_{k_A}\otimes(\kb{0}_B\otimes\kb{e_2}_E + \kb{1}_B\otimes\kb{e_3}).
\]

Now, $B$ will send a qubit in the same state he observed it in and $E$ will attack with $U_R$.  Following this, $A$ will then measure in the $Z$ basis, saving her result in a new register $m_A$.  This results in the following density operator:
\begin{align*}
&\frac{1}{2}\kb{0}_{k_A}\otimes(\kb{00}_{B,m_A}\otimes\kb{e_{0,0}^0} + \kb{01}_{B,m_A}\otimes\kb{e_{0,0}^1} + \kb{10}_{B,m_A}\otimes\kb{e_{1,1}^0} + \kb{11}_{B,m_A}\otimes\kb{e_{1,1}^1})\\
+&\frac{1}{2}\kb{1}_{k_A}\otimes(\kb{00}_{B,m_A}\otimes\kb{e_{0,2}^0} + \kb{01}_{B,m_A}\otimes\kb{e_{0,2}^1} + \kb{10}_{B,m_A}\otimes\kb{e_{1,3}^0} + \kb{11}_{B,m_A}\otimes\kb{e_{1,3}^1}).
\end{align*}

Finally, $A$ and $B$ will only keep this iteration as a key-distillation iteration, if $k_A = m_A$.  Thus, conditioning on this event, the final quantum state, $\rho_{ABE}$, is found to be:
\begin{align}
\rho_{ABE} = &\frac{1}{2N}\kb{0}_A\otimes\left(\kb{0}_B\otimes\kb{e_{0,0}^0} + \kb{1}_B\otimes\kb{e_{1,1}^0}\right)\label{eq:density-op-main}\\
+ &\frac{1}{2N}\kb{1}_A\otimes\left(\kb{1}_B \otimes\kb{e_{1,3}^1} + \kb{0}_B\otimes\kb{e_{0,2}^1}\right).\notag
\end{align}
where we now use the $A$ register to denote $A$'s raw-key bit (the $B$ register is $B$'s raw-key bit), and where $N$ is the following normalization term:
\begin{equation}\label{eq:norm-term}
N = \frac{1}{2}(\bk{e_{0,0}^0} + \bk{e_{1,1}^0} + \bk{e_{1,3}^1} + \bk{e_{0,2}^1}).
\end{equation}

From this, we use Theorem \ref{thm:entropy} to derive the following lower-bound on the conditional entropy $S(A|E)$:
\begin{align}
S(A|E) &\ge \frac{\bk{e_{0,0}^0} + \bk{e_{1,3}^1}}{2N}\left(H\left[\frac{\bk{e_{0,0}^0}}{\bk{e_{0,0}^0} + \bk{e_{1,3}^1}}\right] - H(\lambda_1)\right)\label{eq:entropy}\\
&+ \frac{\bk{e_{1,1}^0} + \bk{e_{0,2}^1}}{2N}\left(H\left[\frac{\bk{e_{1,1}^0}}{\bk{e_{1,1}^0} + \bk{e_{0,2}^1}}\right] - H(\lambda_2)\right),\notag
\end{align}
where:
\begin{align}
\lambda_1 &= \frac{1}{2}\left(1 + \frac{\sqrt{(\bk{e_{0,0}^0} - \bk{e_{1,3}^1})^2 + 4Re^2\braket{e_{0,0}^0|e_{1,3}^1}}}{\bk{e_{0,0}^0} + \bk{e_{1,3}^1}}\right)\label{eq:eigenvalue}\\
\lambda_2 &= \frac{1}{2}\left(1 + \frac{\sqrt{(\bk{e_{1,1}^0} - \bk{e_{0,2}^1})^2 + 4Re^2\braket{e_{1,1}^0|e_{0,2}^1}}}{\bk{e_{1,1}^0} + \bk{e_{0,2}^1}}\right).\notag
\end{align}

Now, let $\pkey_{i,j}$ denote the probability that $A$'s raw-key bit is $i$ and $B$'s raw-key bit is $j$ (conditioning on a key-distillation iteration).  These values are clearly observable by $A$ and $B$; furthermore, from Equation \ref{eq:density-op-main}, they are easily seen to be:
\begin{align*}
\pkey_{0,0} = \frac{\bk{e_{0,0}^0}}{2N} && \pkey_{1,1} = \frac{\bk{e_{1,3}^1}}{2N}\\
\pkey_{0,1} = \frac{\bk{e_{1,1}^0}}{2N} && \pkey_{1,0} = \frac{\bk{e_{0,2}^1}}{2N}
\end{align*}

From this, computing $H(A|B)$ is trivial, namely:
\[
H(A|B) = H(\pkey_{0,0}, \pkey_{0,1}, \pkey_{1,0}, \pkey_{1,1}) - H(\pkey_{0,0} + \pkey_{1,0}).
\]
To compute the key-rate $r$, and also the effective key-rate $\reff$, we need to determine, or bound, those inner-products appearing in the expressions above.  This will be done by looking at various measurement statistics, including mismatched measurements.  When using $\MTWO$, we will rely on results from \cite{krawec2016quantum} (which we summarize in the next sub-section) to determine these bounds.  When using $\MTHREE$, we will consider new measurement statistics and show how they can greatly improve the noise tolerance of the resulting protocol (furthermore, the results we derive when analyzing this three-basis case, can be applied to other two-way protocols and may be of great use to future research in quantum cryptography).

\subsection{Parameter Estimation for $\MTWO$}

When using our protocol in $\MTWO$, we are able to rely on the method of mismatched measurements, for two-way channels from \cite{krawec2016quantum} (which extended results from \cite{QKD-Tom-1,QKD-Tom-2} to two-way semi-quantum protocols) to derive appropriate bounds on the inner-products appearing in the expression for $S(A|E$) derived above.  In this subsection we review these derivations, however, for greater detail the reader is referred to \cite{krawec2016quantum}.

Denote by $\prf_{i,j}$, for $i \in \{0,1,+\}$ and $j \in \{0,1\}$ to be the probability that $B$ measures $\ket{j}$ conditioned on the event $A$ initially sent $\ket{i}$ and that $B$ chose the $\MR$ operation.  Also, denote by $\prr_{i,j,k}$ for $i$ and $j$ as before and $k \in \{0,1,+,-\}$ to be the probability that $A$ observes $\ket{k}$ (when a qubit returns to her) conditioned on the event she initially sent $\ket{i}$ and that $B$ measured (and thus resent) $\ket{j}$ (that is, we condition on the event that $B$ chose operation $\MR$ \emph{and} he actually observed $\ket{j}$).  Finally, denote by $\prr_{i,R,k}$ to be similar, except conditioned on the event $B$ chose to $\R$.  

Note that certain events measure the error in the quantum channel (such as $\prf_{0,1}$) while some measure mismatched measurements (such as $\prf_{+,0}$).  Regardless, these probabilities are all, clearly, observable by users $A$ and $B$ and we will use them to derive bounds on the various inner products appearing in the entropy equations from the previous section.

Later, when we evaluate our resulting key-rate bound (and, also, to compare with BB84+CAD in \cite{BB84-CAD}), we consider a symmetric attack whereby the noise in the quantum channel may be parameterized as follows:
\begin{align}
&\prf_{i,1-i} = Q_F, \text{ for any $i\in\{0,1\}$}\label{eq:qf-sym}\\
&\prr_{i,j,1-j} = Q_R, \text{ for any $i \in \{0,1,+\} ,j \in \{0,1\}$}\notag\\
&\prr_{+,R,-} = Q_X.\notag
\end{align}
Note that this assumption of a symmetric channel, though common in QKD security proofs, is not required in our analysis.  We will derive our key-rate equation in general terms, but also, in parallel, show the symmetric case whereby many simplifications are possible.

From this, we have the following:
\begin{align*}
&\bk{e_{0,0}^0} = \prf_{0,0}\prr_{0,0,0} = (1-Q_F)(1-Q_R)\\
&\bk{e_{1,3}^1} = \prf_{1,1}\prr_{1,1,1} = (1-Q_F)(1-Q_R)\\
&\bk{e_{1,1}^0} = \prf_{0,1}\prr_{0,1,0} = Q_FQ_R\\
&\bk{e_{0,2}^1} = \prf_{1,0}\prr_{1,0,1} = Q_FQ_R \\
&\bk{e_{0,0}^1} = \prf_{0,0}\prr_{0,0,1} = (1-Q_F)Q_R \\
&\bk{e_{1,3}^0} = \prf_{1,1}\prr_{1,1,0} =(1-Q_F)Q_R\\
&\bk{e_{1,1}^1} = \prf_{0,1}\prr_{0,1,1} = Q_F(1-Q_R)\\
&\bk{e_{0,2}^0} = \prf_{1,0}\prr_{1,0,0} = Q_F(1-Q_R)
\end{align*}

With this, along with Equation \ref{eq:norm-term}, we can see that 
\begin{align*}
N &= \frac{1}{2}(\prf_{0,0}\prr_{0,0,0} + \prf_{0,1}\prr_{0,1,0} + \prf_{1,1}\prr_{1,1,1} + \prf_{1,0}\prr_{1,0,1})\\
&= (1-Q_F)(1-Q_R) + Q_FQ_R.
\end{align*}

Consider, now, $\prr_{+,R,-} = Q_X$ (the error in the $X$ basis of \emph{the entire two-way channel}).  It was shown in \cite{krawec2016quantum} that:
\begin{align}
&Q_X = 1-\frac{1}{2}(\Lambda_1 + \Lambda_2 + Re\braket{e_{0,0}^1|e_{1,3}^0} + Re\braket{e_{1,1}^1|e_{0,2}^0} + q_1 + q_2 + \prr_{0,R,+} + \prr_{1,R,+})\notag\\
\Rightarrow & \Lambda_1 +\Lambda_2= 2-2Q_X - (q_1+q_2+\prr_{0,R,+} + \prr_{1,R,+} + Re\braket{e_{0,0}^1|e_{1,3}^0} + Re\braket{e_{1,1}^1|e_{0,2}^0}),\label{eq:L1-2mode}
\end{align}
where $\Lambda_1 = Re\braket{e_{0,0}^0|e_{1,3}^1}$ and $\Lambda_2 = Re\braket{e_{1,1}^0|e_{0,2}^1}$ (note that we chose the notation so that $\Lambda_i$ is the inner product appearing in the expression $\lambda_i$ in Equation \ref{eq:eigenvalue}) and where:
\begin{align}
q_1 &= 2\prf_{+,0}\prr_{+,0,+} - \prf_{+,0} + \prf_{0,0}\left(\frac{1}{2}-\prr_{0,0,+}\right) + \prf_{1,0}\left(\frac{1}{2}-\prr_{1,0,+}\right)\label{eq:q1}\\
q_2 &= 2\prf_{+,1}\prr_{+,1,+} - \prf_{+,1} + \prf_{0,1}\left(\frac{1}{2}-\prr_{0,1,+}\right) + \prf_{1,1}\left(\frac{1}{2}-\prr_{1,1,+}\right)\notag
\end{align}
In fact, it was shown in \cite{krawec2016quantum}, that:
\begin{align}
&Re\braket{e_{0,0}^0|e_{0,2}^1} + Re\braket{e_{0,0}^1|e_{0,2}^0} = q_1\label{eq:q1-b}\\
&Re\braket{e_{1,1}^0|e_{1,3}^1} + Re\braket{e_{1,1}^1|e_{1,3}^0} = q_2\notag,
\end{align}
two equalities which will become important later.

From the Cauchy-Schwarz inequality, we may bound:

\begin{align}
|\braket{e_{0,0}^1|e_{1,3}^0}| &\le \sqrt{\bk{e_{0,0}^1}\bk{e_{1,3}^0}} = \sqrt{\prf_{0,0}\prr_{0,0,1}\prf_{1,1}\prr_{1,1,0}}\label{eq:bound1-asym}\\
|\braket{e_{1,1}^1|e_{0,2}^0}| &\le \sqrt{\bk{e_{1,1}^1}\bk{e_{0,2}^0}} = \sqrt{\prf_{0,1}\prr_{0,1,1}\prf_{1,0}\prr_{1,0,0}}.\notag
\end{align}
Of course, in the symmetric case this simplifies to:
\begin{align}
|\braket{e_{0,0}^1|e_{1,3}^0}| &\le  (1-Q_F)Q_R\label{eq:bound1}\\
|\braket{e_{1,1}^1|e_{0,2}^0}| &\le  Q_F(1-Q_R).\notag
\end{align}

While the above expressions hold for arbitrary attacks, when we restrict ourselves to symmetric attacks, all mismatched events (such as $\prf_{+,0}$ and $\prr_{1,R,+}$) are $1/2$ and so it is easy to see that $q_1 = q_2 = 0$ and thus Equation \ref{eq:L1-2mode} simplifies to:
\begin{equation}\label{eq:L1-2mode-simple}
\Lambda_1 +\Lambda_2 = 1-2Q_X - Re\braket{e_{0,0}^1|e_{1,3}^0} - Re\braket{e_{1,1}^1|e_{0,2}^0}.
\end{equation}

We must actually minimize $S(A|E)$ (Equation \ref{eq:entropy}) to compute the key-rate as we must assume the worst case that Eve chooses an optimal attack, within the above constraints.  To minimize this expression in the symmetric case, it is not difficult to see that we must find the \emph{smallest} $\Lambda_i$ values.  From Equation \ref{eq:bound1} we have:

\begin{equation}\label{eq:l1l2boundm2}
\Lambda_1 +\Lambda_2 \ge 1-2Q_X - (1-Q_F)Q_R - Q_F(1-Q_R).
\end{equation}

When evaluating our key-rate bound using $\MTWO$, we must therefore assume that $\Lambda_1+\Lambda_2$ is in fact equal to the above lower bound and simply optimize over all $\Lambda_2$ satisfying:
\begin{equation}\label{eq:L2-bound}
|\Lambda_2| \le \sqrt{\bk{e_{1,1}^0}\bk{e_{0,2}^1}} = \sqrt{\prf_{0,1}\prr_{0,1,0}\prf_{1,0}\prr_{1,0,1}} = Q_FQ_R,
\end{equation}
(the above follows from the Cauchy-Schwarz inequality).
Note that we cannot simply assume $\Lambda_2 = -Q_FQ_R$ as this may not optimize $H(\lambda_2)$ in Equation \ref{eq:entropy}.

The above can also be applied in the asymmetric case, using Equation \ref{eq:L1-2mode} and Equation \ref{eq:bound1-asym} instead of Equation \ref{eq:L1-2mode-simple}.
\subsection{Parameter Estimation for $\MTHREE$}\label{section:mismatch}

We now turn our attention to our protocol in $\MTHREE$.  While the use of mismatched measurements for  two basis semi-quantum protocols had been developed in \cite{krawec2016quantum}, the use of three bases has never before been considered (either through mismatched measurements nor, to our knowledge, in any SQKD protocol).  Thus we extend those results and introduce this analysis for the first time here; furthermore, we expect these results to extend to future SQKD protocols reliant on three bases and even other fully-quantum protocols reliant on a two-way quantum channel.

When operating in $\MTHREE$, in addition to those statistics analyzed in the previous section, we also are able to incorporate states where $A$ initially sends and/or finally measures in the $Y$ basis.  We first consider the case when $B$ chooses $\R$.  In this case, the iteration is, essentially, a one-way channel with $E$ attacking through the unitary operator $V = U_RU_F$.  We may write the action of $V$ on basis states as follows (recall that Eve starts with her ancilla in the state $\ket{\chi}_E$):
\begin{align*}
V\ket{0,\chi}_{TE} &= \ket{0,g_0} + \ket{1,g_1}\\
V\ket{1,\chi}_{TE} &= \ket{0,g_2} + \ket{1,g_3}.
\end{align*}
where each $\ket{g_i}$ is a linear function of the $\ket{e_{i,j}^k}$ states.  Indeed, it is not difficult to show that:
\begin{align}
\ket{g_0} &= \ket{e_{0,0}^0} + \ket{e_{1,1}^0}\label{eq:g0g3}\\
\ket{g_1} &= \ket{e_{0,0}^1} + \ket{e_{1,1}^1}\notag\\
\ket{g_2} &= \ket{e_{0,2}^0} + \ket{e_{1,3}^0}\notag\\
\ket{g_3} &= \ket{e_{0,2}^1} + \ket{e_{1,3}^1}.\notag
\end{align}

Consider, then, $Q_X = p_{+,R,-}$ which, in this notation, is easily found to be:
\[
Q_X = \frac{1}{2} - \frac{1}{2}Re(\braket{g_0|g_1} + \braket{g_0|g_3} + \braket{g_1|g_2} + \braket{g_2|g_3}).
\]
It was shown in \cite{krawec2016quantum} that, for this ``one-way'' attack (one-way attacks were analyzed in three bases in that prior work), it holds (using our notation here):
\begin{equation}
    Re\braket{g_0|g_3} = 1 - \prr_{+,R,-} - \prr_{0_Y,R,1_Y} - \frac{1}{2}(\prr_{0,R,+} + \prr_{1,R,+} + \prr_{0,R,0_Y} + \prr_{1,R,1_Y} - 2) 
\end{equation}
In the symmetric case, where also $Q_X = \prr_{+,R,-} = \prr_{0_Y,R,1_Y}$, this simplifies to
\begin{equation}\label{eq:g0g3bound1}
Re\braket{g_0|g_3} = 1 - 2Q_X.
\end{equation}
Using Equation \ref{eq:g0g3} to expand $Re\braket{g_0|g_3}$ yields:
\begin{align}
Re\braket{g_0|g_3} &= Re(\braket{e_{0,0}^0|e_{0,2}^1} + \braket{e_{0,0}^0|e_{1,3}^1} + \braket{e_{1,1}^0|e_{0,2}^1} + \braket{e_{1,1}^0|e_{1,3}^1}\notag\\
&= \Lambda_1 + \Lambda_2 + Re\braket{e_{0,0}^0|e_{0,2}^1} + Re\braket{e_{1,1}^0|e_{1,3}^1}.\label{eq:g0g3bound2}
\end{align}

We must compute the real part of the right-most two inner products.  We start by considering $\prr_{i,j,0_Y}$ for $i,j \in \{0,1\}$ (under a symmetric attack these are all $1/2$).  Consider, first, $\prr_{0,0,0_Y}$.  Tracing the evolution of the qubit as it travels through Eve's lab in the forward channel, is measured by $B$ and observed to be $0$ (i.e., we are conditioning on the outcome being $\ket{0}$ for this statistic), attacked by $E$ again, and finally returning to $A$.  The state is found to be:
\begin{align*}
\ket{0} &\mapsto \ket{0,e_0} + \ket{1,e_1} \mapsto \frac{\ket{0,e_0}}{\sqrt{\prf_{0,0}}}\\
&\mapsto \frac{\ket{0,e_{0,0}^0} + \ket{1,e_{0,0}^1}}{\sqrt{\prf_{0,0}}}
= \frac{\ket{0_Y}(\ket{e_{0,0}^0} - i\ket{e_{0,0}^1}) + \ket{1_Y}(\ket{e_{0,0}^0} + i\ket{e_{0,0}^1})}{\sqrt{2\prf_{0,0}}},
\end{align*}
from which it is clear that:
\[
\prr_{0,0,0_Y} = \frac{1}{2} + \frac{Im\braket{e_{0,0}^0|e_{0,0}^1}}{\prf_{0,0}}.
\]
(Note that, above, we took advantage of the fact that $\prf_{0,0} = \bk{e_0} = \bk{e_{0,0}^0} + \bk{e_{0,0}^1}$ due to unitarity of $U_R$).
This allows $A$ and $B$ to learn the imaginary part of $\braket{e_{0,0}^0|e_{0,0}^1}$.  Similarly, the following identities may be derived:
\begin{align}
Im\braket{e_{0,0}^0|e_{0,0}^1} &= \prf_{0,0}\left(\prr_{0,0,0_Y} - \frac{1}{2}\right)\label{eq:im:00Y}\\
Im\braket{e_{0,2}^0|e_{0,2}^1} &= \prf_{1,0}\left(\prr_{1,0,0_Y} - \frac{1}{2}\right)\notag\\
Im\braket{e_{1,1}^0|e_{1,1}^1} &= \prf_{0,1}\left(\prr_{0,1,0_Y} - \frac{1}{2}\right)\notag\\
Im\braket{e_{1,3}^0|e_{1,3}^1} &= \prf_{1,1}\left(\prr_{1,1,0_Y} - \frac{1}{2}\right)\notag
\end{align}

In the asymmetric case, we can use these identities in our calculations. In a symmetric attack, the above identities are all equal to zero. These identities will become important momentarily.

Next, we consider statistics of the form $\prr_{0_Y,j,0_Y}$ for $j\in \{0,1\}$ (again, under a symmetric attack, these should be $1/2$, and in arbitrary attacks they would be observed values).  First, consider the case when $j=0$.  Tracing the evolution of the qubit in this instance, we find:
\begin{align}
\ket{0_Y} &\mapsto \frac{1}{\sqrt{2}}(\ket{0}(\ket{e_0} + i\ket{e_2}) + \ket{1}(\ket{e_1} + i\ket{e_3}))
\mapsto \frac{\ket{0}(\ket{e_0} + i\ket{e_2})}{\sqrt{2\prf_{0_Y,0}}}\\
&\mapsto \frac{\ket{0}(\ket{e_{0,0}^0} + i\ket{e_{0,2}^0}) + \ket{1}(\ket{e_{0,0}^1} + i\ket{e_{0,2}^1})}{\sqrt{2\prf_{0_Y,0}}}\label{eq:Y0Yp2}\\
&= \frac{\ket{0_Y}(\ket{e_{0,0}^0} - i\ket{e_{0,0}^1} + i\ket{e_{0,2}^0} + \ket{e_{0,2}^1}) + \ket{1_Y}(\ket{e_{0,0}^0} + i\ket{e_{0,0}^1} + i\ket{e_{0,2}^0} - \ket{e_{0,2}^1})}{\sqrt{4\prf_{0_Y,0}}}.
\end{align}
Using the above, along with properties of unitarity of $U_F$ and $U_R$ (in particular, that $\bk{e_{0,0}^0} + \bk{e_{0,0}^1} = \bk{e_0} = \prf_{0,0}$ and $\bk{e_{0,2}^0} + \bk{e_{0,2}^1} = \bk{e_2} = \prf_{1,0}$), we find:
\begin{align}
\prr_{0_Y,0,0_Y} &= \frac{1}{4\prf_{0_Y,0}}[\prf_{0,0} + \prf_{1,0} + 2(Re\braket{e_{0,0}^0|e_{0,2}^1} - Re\braket{e_{0,0}^1|e_{0,2}^0})\label{eq:Y0Y}\\
&+ 2(Im\braket{e_{0,0}^0|e_{0,0}^1} - Im\braket{e_{0,0}^0|e_{0,2}^0})\notag\\
&+ 2(Im\braket{e_{0,2}^0|e_{0,2}^1} - Im\braket{e_{0,0}^1|e_{0,2}^1})]\notag
\end{align}
As discussed earlier, $Im\braket{e_{0,0}^0|e_{0,0}^1}$ and $Im\braket{e_{0,2}^0|e_{0,2}^1}$ can be determined through observable parameters (see Equation \ref{eq:im:00Y}).  The remaining two imaginary parts in the above expression can also be observed by considering $\prr_{0_Y,0,0}$ and $\prr_{0_Y,0,1}$.  Indeed, from Equation \ref{eq:Y0Yp2}, we easily find:
\begin{align*}
\prr_{0_Y,0,0} = \frac{1}{2\prf_{0_Y,0}}(\bk{e_{0,0}^0} + \bk{e_{0,2}^0} - 2Im\braket{e_{0,0}^0|e_{0,2}^0})
\end{align*}
Noting that $\bk{e_{0,0}^0} = \prf_{0,0}\prr_{0,0,0}$ and $\bk{e_{0,2}^0} = \prf_{1,0}\prr_{1,0,0}$, we find:
\[
Im\braket{e_{0,0}^0|e_{0,2}^0} = \frac{1}{2}(\prf_{0,0}\prr_{0,0,0} + \prf_{1,0}\prr_{1,0,0}) - \prf_{0_Y,0}\prr_{0_Y,0,0}.
\]
Similarly, it is easy to see that:
\[
Im\braket{e_{0,0}^1|e_{0,2}^1} = \frac{1}{2}(\prf_{0,0}\prr_{0,0,1} + \prf_{1,0}\prr_{1,0,1}) - \prf_{0_Y,0}\prr_{0_Y,0,1}.
\]

At this point, we may use Equation \ref{eq:Y0Y} to determine the difference $Re\braket{e_{0,0}^0|e_{0,2}^1} - Re\braket{e_{0,0}^1|e_{0,2}^0}$ (all other inner-products appearing in that equation may be observed).  Using Equations \ref{eq:q1} and \ref{eq:q1-b}, we also determine the sum $Re\braket{e_{0,0}^0|e_{0,2}^1} + Re\braket{e_{0,0}^1|e_{0,2}^0}$ allowing one to determine $Re\braket{e_{0,0}^0|e_{0,2}^1}$ as functions of observed statistics.

In the event of a symmetric attack, it is not difficult to show that, in fact, we have:
\[
Re\braket{e_{0,0}^0|e_{0,2}^1} - Re\braket{e_{0,0}^1|e_{0,2}^0} = 0.
\]
Since $q_1$ (from Equations \ref{eq:q1} and \ref{eq:q1-b}) is also $0$ in this event, we have:
\[
Re\braket{e_{0,0}^0|e_{0,2}^1} + Re\braket{e_{0,0}^1|e_{0,2}^0} = 0.
\]
These two equations, combined, imply that the real part of each inner product individually is, in fact, zero (most importantly, $Re\braket{e_{0,0}^0|e_{0,2}^1} = 0$).  By repeating the above process for $\prr_{0_Y,1,0_Y}$ (along with $\prr_{0_Y,1,1}$ and $\prr_{0_Y,1,0}$ and $q_2$), we also determine $Re\braket{e_{1,1}^0|e_{1,3}^1}$ (and, again, in the symmetric case, it evaluates to $0$).  Combining this, then, with Equations \ref{eq:g0g3bound1} and \ref{eq:g0g3bound2}, we find (in the symmetric case):
\begin{equation}\label{eq:L1L2-M3}
\Lambda_1 + \Lambda_2 = 1-2Q_X.
\end{equation}

For asymmetric channels we can substitute the potentially non-zero values of $Re\braket{e_{1,1}^0|e_{1,3}^1}$ and $Re\braket{e_{0,0}^0|e_{0,2}^1}$ (found using the method described above) into Equations \ref{eq:g0g3bound1}, \ref{eq:g0g3bound2}.

Note this is significantly improved over the bound attained when only considering two bases (Equation \ref{eq:l1l2boundm2}).  To evaluate $S(A|E)$ in $\MTHREE$, we must simply minimize the above expression over all $\Lambda_2$ bounded by Equation \ref{eq:L2-bound}.

\subsection{Summary}
We now summarize our security analysis and how to actually apply it.  The main computation is our bound on $S(A|E)$ in Equation \ref{eq:entropy}.  Many of the inner-products appearing in that expression, such as $\bk{e_{0,0}^0}$ may be directly observed based on the observed channel noise $Q_F$ and $Q_R$ as discussed in Section III.A.  The inner-products appearing in the $\lambda_i$ expressions, however, must be bounded based on mismatched measurements.  Here, if one is using $\MTWO$, the results in Section III.A may be used and, in particular, Equations \ref{eq:l1l2boundm2} and \ref{eq:L2-bound}.  If $\MTHREE$ is used, then additional statistics may be gathered allowing one to get tighter bounds on these expressions leading to Equation \ref{eq:L1L2-M3}.  Either way, one has a single free parameter, $\Lambda_2$, bounded by Equation \ref{eq:L2-bound} (a function of the channel noise).  From this, one may minimize Equation \ref{eq:entropy}, over all $\Lambda_2$ in this range (note that $\Lambda_i$ is the inner-product appearing in $\lambda_i$ needed to compute Equation \ref{eq:entropy}; in particular, $\Lambda_2 = Re\braket{e_{1,1}^0|e_{0,2}^1}$).  One takes the minimum as we must assume the worst case in that $E$'s attack gives her the most information while still conforming to those restrictions placed on the observed noise and mismatched statistics.

\section{EVALUATION}\label{section:eval}

While our key-rate equations apply to any arbitrary channel, to actually evaluate our bounds and compare with other protocols, we consider two forms of attacks in particular: \textbf{independent} and \textbf{dependent} channels.  Both cases are symmetric attacks and parameterized by two values $Q$, the $Z$ basis error in one channel (i.e., $Q_F = Q_R = Q$) and $Q_X$ the $X$ (and $Y$) basis error observed when the qubit travels through both channels (when $B$ reflects); i.e., $Q_X = p_{+,R,-} = p_{0_Y,R,1_Y}$. For the independent case we have $Q_X = 2Q(1-Q)$ (i.e., an error in the reflection case, where a qubit has to travel through both channels, occurs if the qubit flips in the forward channel but not the reverse or it flips in the reverse channel but not the forward).  The second case, the dependent channel, we take to mean $Q_X = Q$.  These two channels are commonly evaluated when discussing security and noise tolerances of two-way protocols in general \cite{QKD-TwoWaySecure,krawec2016quantum} and so we consider them here.  We stress, however, that our key-rate bound is applicable to any observed quantum noise (indeed, the equations derived in the previous section apply to any observed parameters).

\begin{figure}
  \centering
  \includegraphics[width=300pt]{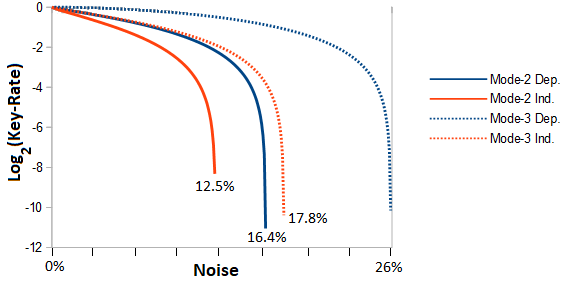}
  \caption{Showing the key-rate of our protocol in both modes of operation and under both channel assumptions.  See text for more details.  Color online; Blue-Dashed line (top-most dashed line reaching $26\%$) is $\MTHREE$ Dep.; Red-Dashed line (lower most of the two dashed lines reaching $17.8\%$) is $\MTHREE$ Ind.; Blue-Solid line (top most solid line reaching $16.4\%$) is $\MTWO$ Dep.; finally, Red-Solid line (lower-most solid line reaching $12.5\%$) is $\MTWO$ Ind. Horizontal axis is the noise, $Q$, in the channel (the probability that a state depolarizes in one channel) while vertical axis is the log of the key-rate (the ratio of secure secret key bits to the raw-key size).}\label{fig:key-rate}
\end{figure}

\begin{figure}
  \centering
  \includegraphics[width=300pt]{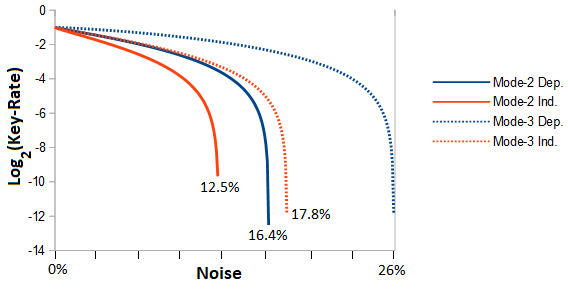}
  \caption{Showing the \emph{effective} key-rate of our protocol in both modes of operation and under both channel assumptions.  This effective key-rate is lower, though of course noise tolerance is unaffected.  See text for more details. Color online; Blue-Dashed line (top-most dashed line reaching $26\%$) is $\MTHREE$ Dep.; Red-Dashed line (lower most of the two dashed lines reaching $17.8\%$) is $\MTHREE$ Ind.; Blue-Solid line (top most solid line reaching $16.4\%$) is $\MTWO$ Dep.; Finally, Red-Solid line (lower-most solid line reaching $12.5\%$) is $\MTWO$ Ind. Horizontal axis is the noise, $Q$, in the channel (the probability that a state depolarizes in one channel) while vertical axis is the log of the key-rate (the ratio of secure secret key bits to the total number of qubits sent).}\label{fig:eff-keyrate}
\end{figure}

We evaluate our key-rate bound under these two assumptions looking for the maximal $Q$ for which $r > 0$; these results are shown in Table \ref{tbl:noise-tol}.  This noise tolerance is substantially higher, in both cases, than any prior SQKD protocol which currently has a noise tolerance analysis.  This is also higher than other fully-quantum protocols using two-way channels for which noise tolerance bounds are currently known.  We also plot the key-rate and the effective key-rate in Figures \ref{fig:key-rate} and \ref{fig:eff-keyrate}.  For the effective key-rate, we use Equation \ref{eq:eff-keyrate} with $p_{acc} = (1-Q)^2+Q^2$.

\begin{table}
\centering
\begin{tabular}{|r|c|c|}
\hline
Protocol&Independent: $Q_X = 2Q(1-Q)$ & Dependent: $Q_X = Q$\\
\hline
$\MTWO$ & $12.5\%$ & $16.4\%$\\
$\MTHREE$ & $17.8\%$ & $26.0\%$\\
\hline
Original SQKD \cite{SQKD-first,krawec2016quantum} & $7.9\%$ & $11\%$\\
Single-State SQKD \cite{zhang2018security} & $7.55\%$ & $9.65\%$\\
Reflection SQKD \cite{Krawec2016Reflect} & $3.64\%$ & $5.36\%$\\
\hline
LM05 \cite{QKD-LM05,QKD-TwoWaySecure} & $6\%$ & $11\%$\\
SDC-PP \cite{QKD-PingPong,QKD-TwoWaySecure} & $8.5\%$ & $12\%$\\
\hline
BB84-CAD \cite{CAD,BB84-CAD,chau2002practical} & \multicolumn{2}{|c|}{$27.6\%$}\\
\hline
\end{tabular}
\caption{Maximal noise tolerance of our SQKD protocol under both dependent and independent channel scenarios.  We also compare with three other SQKD protocols which, as of writing, have noise tolerance bounds, and two fully-quantum protocols which require a two-way quantum channel, the LM05 protocol \cite{QKD-LM05} and a variant of the PingPong protocol (introduced in \cite{QKD-PingPong}, though the variant is from \cite{QKD-TwoWaySecure}).  Noise tolerances for these last two protocols come from \cite{QKD-TwoWaySecure} which, to our knowledge, represents the most current best-known bounds for these protocols.  In all cases, our semi-quantum protocol's noise tolerance is significantly higher.  We note that, the techniques we used to attain this high noise tolerance (namely, utilizing the reverse quantum channel strategically) could probably be adapted to LM05 and SDC-PP improving their noise tolerances also.  The adaption is not straight-forward, however, due to the manner in which these protocols distill a raw key and so we leave this investigation as future work. Finally, we compare with the fully quantum BB84 using CAD from \cite{CAD,BB84-CAD} which has a superior noise-tolerance ability.}\label{tbl:noise-tol}
\end{table}

\subsection{Comparison to BB84 with CAD}

Considering the positive results in the previous section, it is useful to compare to other fully-quantum protocols and also to attempt to discover exactly \emph{why} the noise tolerance is so high in our protocol.  Note that, since $A$ is rejecting all iterations where she does not receive the same $Z$ basis qubit state she initially sent, this serves to greatly reduce the raw-key error.  That is, $A$ and $B$ can expect to end the protocol with a raw-key that is more highly correlated than previous SQKD protocols (thus reducing the information leaked during error correction).  However, such a process can also be done classically to, say, BB84 through the use of \emph{classical advantage distillation} (CAD).  In this section we compare our protocol to BB84 with CAD to show similarities, but also interesting quantum-level differences.

Recall the BB84 protocol (including the six-state BB84), the quantum communication stage of which consists of the following process:

$ $\newline
\textbf{BB84} \cite{QKD-BB84}:
$ $\newline
\textbf{Public Constants: }
\begin{itemize}
  \item $\texttt{Mode} = \texttt{XZ}$ or $\texttt{XYZ}$ specifying whether to use two bases ($Z$ and $X$) or three ($Z$, $X$, and $Y$).  Furthermore, we consider the asymmetric version of BB84 \cite{QKD-renner-keyrate} whereby only the $Z$ basis is used for key distillation while any other basis is used only for error testing.
  \item $p \in (0,1)$, the probability of choosing the $Z$ basis in a particular iteration.
\end{itemize}
$ $\newline
\textbf{Quantum Communication Stage:}
This stage repeats the following process until a sufficiently large raw-key is produced.
\begin{enumerate}
  \item \textbf{Alice's Preparation:}
  \begin{itemize}
    \item With probability $p$, $A$ sets an internal private register to $b_A = Z$; otherwise, with probability $1-p$, she sets $b_A$ to be $X$ if $\texttt{Mode} = \texttt{XZ}$ or she sets $b_A$ to be $X$ or $Y$ (with probability $(1-p)/2$ each) if $\texttt{Mode} = \texttt{XYZ}$.
    \item If $b_A = Z$, then $A$ sets an internal private register $k_A$ to be $0$ or $1$ with uniform probability and she sends the computational qubit state $\ket{k_A}$ to $B$.
    \item Otherwise, if $b_A = X$, $A$ sends the state $\ket{+}$ or $\ket{-}$ to $B$ (choosing uniformly at random and, of course, saving her choice in a private register); or, if $b_A = Y$ (which only happens when $\texttt{Mode} = \texttt{XYZ}$), she sends the state $\ket{0_Y}$ or $\ket{1_Y}$ to $B$ (again, choosing uniformly at random).
  \end{itemize}

  \item \textbf{Bob's Operation:}
  \begin{itemize}
    \item When $B$ receives a qubit he will set an internal private register $b_B$ to $Z$ with probability $p$; otherwise, with probability $1-p$, if $\texttt{Mode} = \texttt{XZ}$, he will set $b_B$ to be $X$ or, if $\texttt{Mode} = \texttt{XYZ}$, he will set it to be $X$ or $Y$ (with probability $(1-p)/2$ each).
    \item Bob will measure the qubit in the basis specified by $b_B$, saving the result.
  \end{itemize}

  \item \textbf{Communication:} \emph{(Note this step may be performed after performing the above steps for sufficiently long.)}
  \begin{itemize}
    \item Using the authenticated channel, $A$ discloses $b_A$ and $B$ discloses $b_B$.  For any iteration where $b_A = b_B = Z$, they will save their preparation (for $A$) and measurement (for $B$) results to use as their raw key.  For all other iterations, $A$ and $B$ will disclose complete information on their choices and measurement outcomes for use in error checking.
    \item $A$ will chose a random subset of the raw key for use in error checking - namely, for any iteration in this subset, $A$ and $B$ will disclose their raw key result on this subset (and, of course, remove these results from their raw key).
  \end{itemize}
\end{enumerate}

Note that, as shown in \cite{QKD-BB84-Modification}, in the asymptotic scenario, we may set $p$ arbitrarily close to $1$ to improve efficiency of the protocol.

Following the execution of this protocol, before further processing the raw-key through error correction and privacy amplification, users may choose to perform a \emph{Classical Advantage Distillation} (CAD) protocol. This is a two-way communication process using the authenticated classical channel, which attempts to create an additional advantage for $A$ and $B$ over $E$ by processing the raw key and producing a new, shorter, raw key. This shorter key is then, subsequently processed further with error correction and privacy amplification.  We will consider the following CAD protocol, discussed in detail in \cite{CAD,BB84-CAD}, which is parameterized by a block-length parameter $C \ge 1$:
\begin{enumerate}
  \item $A$ will select a group of raw-key bits of size $C$ such that all $C$ bits are the same value (all will be $0$ or all will be $1$).  $A$ then sends the \emph{indices} of these bits to $B$ (of course, she keeps their value secret).
  \item $B$ will check his raw key to see if all $C$ bits are the same value on his end.  If they are, he will tell $A$ to ``accept'' this block; otherwise to ``reject.''  (Note that all communication in this step and the previous are done using the authenticated classical channel.)
  \item If $B$ accepts, both parties compress the $C$ equal-valued bits to a single bit in the natural way; this new bit is added to the new raw-key while the old block of $C$ bits is discarded from the original raw-key.  If $B$ rejects, both parties discard all $C$ bits.
\end{enumerate}
We denote by BB84-XZ$[C]$ to mean four-state BB84 using CAD with a block size of $C$; similarly, we use BB84-XYZ$[C]$ to mean the same for the six-state BB84.

Since we are comparing to a SQKD protocol using a two-way quantum channel, we will assume, when running BB84, that users also have access to a two-way quantum channel on which they will run two independent copies of BB84.  That is, they will run BB84 with CAD on the forward channel resulting in a secret key of size $\ell(N_F)$ and they will run, independently, BB84 with CAD on the reverse channel resulting in a secret key of size $\ell(N_R)$.  They will then combine the two keys resulting in a secret key of size $\ell(N_F)+\ell(N_R)$ (thus, the key-rate will double).  This allows one to better compare two-way protocols to one-way protocols and was also the evaluation method used in \cite{QKD-TwoWaySecure} when they compared LM05 and SDC-PP (which, like our SQKD protocol, require a two-way quantum channel) to BB84.

On a single instance of BB84 with CAD, observe that, even in the noise-less case, the new raw-key will shrink by a factor of $1/C$ (i.e., if the original raw-key before CAD was $M$ bits long, the new raw-key will be only $M/C$ bits long).  In the event the channel is noisy, this loss of raw-key material may be even more extreme.  However, the advantage to this process is that the new raw-key should be more highly correlated than the previous one (indeed, for an error to exist in the new raw key, all $C$ bits on $B$'s end must be wrong - an event which occurs with probability, roughly, $Q^C$).

In \cite{BB84-CAD}, the BB84 protocol (both the four and six state versions) were analyzed with this particular CAD process.  In that source, they computed the key-rate to be:
\begin{equation}\label{eq:rate-bb84}
r_{BB84}(C) = 1 - h(e_C) - (1-e_C)h\left(\frac{1-\Lambda_{eq}^C}{2}\right) - e_C h\left(\frac{1-\Lambda_{diff}^C}{2}\right),
\end{equation}
where:
\[
e_C = \frac{Q^C}{Q^C + (1-Q)^C},
\]
is the error in the new raw-key (after applying CAD) - this is easy to compute as the error will be the probability that all $C$ bits are wrong on $B$'s end (with probability $Q^C$) conditioned on the probability that CAD ``accepts'' the block (which happens only if they are all equal or all different, thus this occurs with probability $Q^C + (1-Q)^C$).

The eigenvalues $\Lambda_{eq}$ and $\Lambda_{diff}$ depend on whether one is considering the four or six state protocol.  Using here also results in \cite{BB84-CAD}, we have:
\begin{align*}
\left.
\begin{array}{l}
\Lambda_{eq} = \frac{1-2Q}{1-Q}\\\\
\Lambda_{diff} = 0
\end{array}
\right \} &\text{Six-State BB84}\\\\
\left.
\begin{array}{l}
\Lambda_{eq} = \frac{1-3Q+2\lambda_4}{1-Q}\\\\
\Lambda_{diff} = \frac{|Q-2\lambda_4|}{Q}
\end{array}
\right \} &\text{Four-State BB84}
\end{align*}
where, for four-state BB84, we must optimize over all $\lambda_4 \in [0,Q]$.

Returning to our protocol, we note that there is a similarity to running BB84 with CAD, setting $C=2$.  Of course, to perform a fair comparison, we must also consider the effective key-rate of both protocols.  For our SQKD protocol, we use Equation \ref{eq:eff-keyrate} where we have $p_{acc} = (1-Q)^2 + Q^2$.

For BB84 using CAD with a block size of $C$, it is not difficult to see that:
\[
N = K \cdot p_{acc} \cdot \frac{1}{C},
\]
where $N$ is the expected raw-key size, $K$ is the number of qubits prepared in total, and $p_{acc} = (1-Q)^C + Q^C$.  Thus:
\[
\reff_{BB84[C]} = \frac{\ell(N)}{K} = \frac{1}{C}\cdot p_{acc}\cdot r_{BB84}(C).
\]
Of course, when running two independent instances of BB84 (one instance per quantum channel), these key-rates double.

\begin{figure}
  \centering
  \includegraphics[width=350pt]{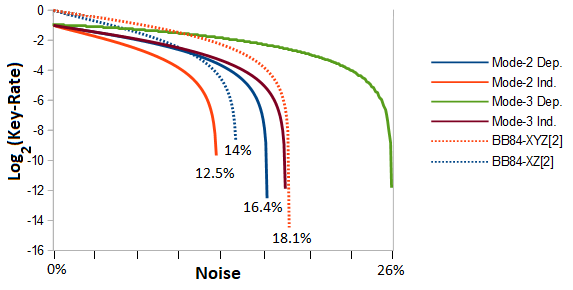}
  \caption{Showing the effective key-rate of our protocol, compared with the effective key-rate of \emph{two instances} of BB84 with CAD applied using a block size of 2. Color available online; Solid-Green line (top-most of the solid lines, reaching $26\%$) is $\MTHREE$ Dep.; Solid-Magenta (solid line reaching near $18\%$) is $\MTHREE$ Ind.; Solid Blue (solid line reaching $16.4\%$) is $\MTWO$ Dep.; Solid-Red (lowest line, reaching $12.5\%$) is $\MTWO$ Ind.; Dashed-Red (highest dashed line reaching $18.1\%$) is BB84-XYZ$[2]$; finally, Dashed-Blue (lowest of the dashed lines, reaching $14\%$) is BB84-XZ$[2]$. Horizontal axis is the noise, $Q$, in the channel (the probability that a state depolarizes in one channel) while vertical axis is the log of the key-rate (the ratio of secure secret key bits to the number of qubits sent in total).}\label{fig:keyrate-graph2}
\end{figure}

A graph of the resulting key-rates, assuming two independent instances of BB84 is run as discussed earlier, is shown in Figure \ref{fig:keyrate-graph2}.  Here we notice that the six-state BB84 almost exactly agrees with our protocol in $\MTHREE$ for the independent channel case although BB84 is better.  There is a striking improvement in key-rate in the dependent case (especially for $\MTHREE$).  Two instances of BB84-XY[2] is most comparable to our protocol in $\MTWO$; similarly, BB84-XYZ[2] is most comparable to $\MTHREE$.

\begin{figure}
  \centering
  \includegraphics[width=350pt]{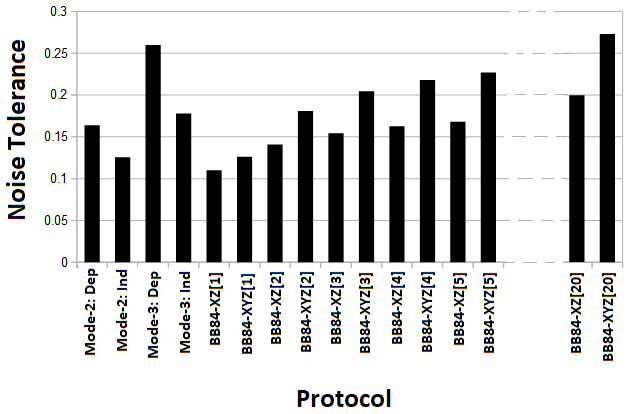}
  \caption{Showing the noise tolerance for our protocol and two instances of BB84 with various CAD block sizes.  See text for further explanations and observations.  Vertical axis is maximal noise tolerance where noise, $Q$, is the probability that a state depolarizes in one channel.}\label{fig:noise-tol}
\end{figure}

In the independent case, two copies of BB84 provide a superior key-rate and noise tolerance; in the dependent case, our protocol produces a better noise tolerance (and often better key-rate).  This is due to the fact that our protocol is able to take advantage of quantum noise that may ``reverse'' itself when a qubit travels back whereas two copies of BB84 remain independent.  Note that this is a realistic possibility for some fiber channels where any phase error picked up in the forward direction is ``undone'' in the reverse \cite{QKD-TwoWaySecure,lucamarini2014quantum}.  Thus, we summarize that if such a channel were implemented in practice, our SQKD protocol (or perhaps another fully-quantum two way protocol augmented with the techniques we developed in this paper) may be an excellent candidate to improve secure communication.  However, even in the independent case, we achieve a very similar noise tolerance and effective key-rate to BB84 with CAD, without having to perform a two-way CAD process over the authenticated channel (though BB84 is better in the independent case). A summary comparison of noise tolerances is shown in Figure \ref{fig:noise-tol}.

Note that, as the CAD block size approaches infinity, the BB84 protocol can tolerate up to $27.6\%$ noise \cite{chau2002practical,BB84-CAD}, surpassing our protocol's performance in all modes and evaluated channels.



To look at this further, let us compare not the final key-rate of our protocol to BB84 with CAD, but instead only $E$'s uncertainty as measured by $S(A|E)$ on a raw-key iteration.  Rather surprisingly, we show that $E$'s uncertainty is greater in our case than in the case of BB84 with CAD for certain channels.  Again, using computations from \cite{BB84-CAD}, and the definition of quantum mutual information: $I(A:E) = S(A) + S(E) - S(AE) = S(A) - S(A|E) = 1 - S(A|E)$ (since we are assuming a symmetric attack in both protocols thus $A$'s key bit is unbiased and, so, equally likely to be $0$ or $1$ yielding $S(A) = 1$).  In this case, it is trivial algebra to show (using Equation \ref{eq:rate-bb84} and basic definitions of mutual information):
\begin{equation}
S(A|E)_{BB84[C]} = 1 - (1-e_C)h\left(\frac{1-\Lambda_{eq}^C}{2}\right) - e_C h\left(\frac{1-\Lambda_{diff}^C}{2}\right).
\end{equation}

\begin{figure}
  \centering
  \includegraphics[width=375pt]{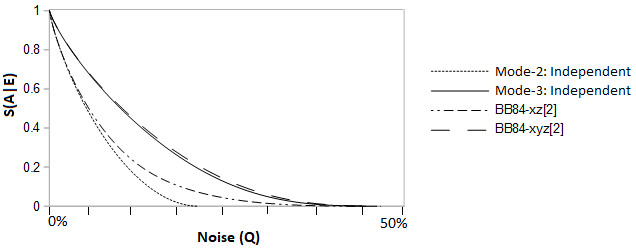}
  \caption{Showing $E$'s uncertainty, as measured by $S(A|E)$ of our protocol, along with BB84 with a CAD block size of $2$.  For our two-way protocol, we are assuming an independent channel in this figure.  That is, for our SQKD protocol, the $X$ basis error rate is almost twice what it is when evaluating BB84 in this graph.  Horizontal axis is noise (the probability that a state depolarizes in one channel) while vertical axis is the conditional entropy of the system.}\label{fig:entropy-ind}
\end{figure}

\begin{figure}
  \centering
  \includegraphics[width=375pt]{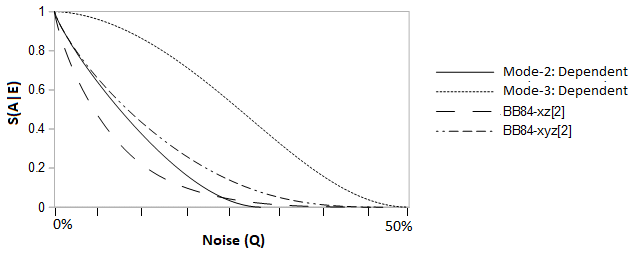}
  \caption{Showing $E$'s uncertainty, as measured by $S(A|E)$ of our protocol, along with BB84 with a CAD block size of $2$.  For our two-way protocol, we are assuming a dependent channel in this figure.  Higher uncertainty is better for $A$ and $B$. Horizontal axis is noise (the probability that a state depolarizes in one channel) while vertical axis is the conditional entropy of the system.}\label{fig:entropy-dep}
\end{figure}

Figure \ref{fig:entropy-ind} shows a comparison of $S(A|E)$ for our protocol and BB84 with CAD in the independent case; Figure \ref{fig:entropy-dep} shows the same but for the dependent channel case.  For the dependent case, Eve's uncertainty is far greater than BB84[2] in all cases except for $\MTWO$ when $Q > 25\%$.  For the independent case, the uncertainty $E$ has on $A$'s raw key bit is almost identical between $\MTHREE$ and BB84-XYZ[2].  But for $\MTWO$ and BB84-XZ[2], the latter admits greater uncertainty for Eve.  Note that, in these computations, we are looking at only one copy of BB84 to compare uncertainty on a single instance and not overall key-rate.

\section{CHANNEL LOSS}\label{section:loss}

While our main focus has been the analysis of our protocol under a loss-less channel, in this section we provide a preliminary analysis of our protocol's behavior on a channel with loss.  In particular, this will allow us to compute bounds on the maximal distance over which a key may be successfully distilled.  To do this analysis, we must first specify how $A$ and $B$ should behave when a loss event occurs (i.e., when either or both $A$ and $B$ detect a vacuum) on any particular iteration.  Due to the drastic increase in the probability of a vacuum over distance when running on fiber channels \cite{QKD-survey}, we will attempt to increase the efficiency of the protocol by assuming that $A$ will not discard iterations where she detects a vacuum on the return path.  If $B$ detects a vacuum, he will discard the iteration.  Thus, considering our protocol description from Section II, we augment it with the following rules: On step 2, Bob's Operation, if $B$ receives a vacuum after choosing $\MR$, he will simply inform $A$ to discard the iteration (this may be done later of course).  On step 3, Alice's Measurement, we add the rule that if $b_A = b'_A = Z$ and if she detects a vacuum, she sets her internal register $\texttt{Accept} = 1$.  That is, she will assume the measurement would have worked out.  This gives $E$ an advantage, and more strategic choices of protocol implementation would be worth considering for future work.  However, here, we are only interested in getting a rough bound on the tolerance to loss of our protocol.

Now, to analyze the protocol, we will assume the worst case, that on any vacuum event, $E$ actually gains \emph{full information} on $A$'s key-bit register.  This is obviously a strong assumption in favor of the adversary.  In real life, $E$ would mostly likely have some uncertainty.  Future work may be able to tighten and improve on this analysis.

To perform the required analysis, we will model the forward and reverse channels as symmetric channels.  In the forward direction, this maps a density operator $\rho$ to:
\[
\mathcal{E}_F(\rho) = (1-p_l)U_F\rho U_F^* + p_l\kb{vac}_T\otimes\mu_E,
\]
where $p_l$ is the probability of a loss in one channel (forward or backward); $\ket{vac}_T$ is an orthonormal basis state in the transit Hilbert space modeling a vacuum event; $U_F$ is $E$'s attack operator used in the protocol analysis earlier; and $\mu_E$ is some arbitrary density operator modeling $E$'s state but giving her full information on $A$'s raw key bit (again, an unrealistic assumption giving great advantage to the adversary - in practice, the bound we derive can only be better).  The reverse channel is similar, using $U_R$ instead of $U_F$.  Of course we must augment $U_R$ to act on $\ket{vac}$, however this action may be arbitrary since $B$ will eventually reject the iteration anyway.  One may think of this attack model as Eve choosing to drop a photon randomly (and somehow gaining full information on $A$'s key in the process, a strong assumption in favor of the adversary) or letting it travel but probing it with operator $U_F$/$U_R$.

Using this attack model, and realizing that $A$ always accepts when she detects a vacuum, we may model a single iteration of the protocol, conditioning on a raw key being distilled, as follows.  In the forward channel, the operator, after $B$'s operation but before sending to $E$ again, is of the form:
\[
(1-p_l)\mu_{ABE} + p_l\mu_{discard},
\]
where $\mu_{ABE}$ is the operator identical to the previous no-loss case, and $\mu_{discard}$ is some arbitrary operator where $B$'s flag specifies that he will discard this iteration (as he detected a vacuum); nothing Eve does to this operator will later matter as $B$ will always send the message to discard that iteration.  On the reverse channel, after $A$'s final measurement, but before $B$ informs whether he received a vacuum or not, this system evolves to:
\[
(1-p_l)^2\rho_{ABE} + (1-p_l)p_l \tau_{ABE} + \mu'_{discard},
\]
where $\rho_{ABE}$ is the density operator derived in the previous section (where there was no loss), see Equation \ref{eq:density-op-main} and $\tau_{ABE}$ is some arbitrary density operator where $E$ has full information, yet $A$ and $B$ eventually distill a key-bit.  Finally, $\mu'_{discard}$ is some arbitrary, non-normalized, operator on which $B$ will eventually report ``reject.''  $B$ will then inform $A$ to discard the iteration if he initially detected a vacuum (i.e., we will project out the $\mu'_{discard}$ case) yielding the following density operator:
\begin{equation}
\sigma_{ABE} = \frac{(1-p_l)^2\rho_{ABE} + (1-p_l)p_l \tau_{ABE}}{(1-p_l)^2 + (1-p_l)p_l} = (1-p_l)\rho_{ABE} + p_l\tau_{ABE}.
\end{equation}
Using the concavity of conditional von Neumann entropy, and the fact that $E$ has no uncertainty on vacuum events in this scenario, we have:
\[
S(A|E)_\sigma \ge (1-p_l)S(A|E)_\rho,
\]
and $S(A|E)_{\rho}$ may be bounded as before (though, of course, all measured statistics used must now be taken to mean conditioned on no loss - since we are assuming still the asymptotic scenario, loss will not affect statistic gathering so long as the probability of a photon being received is non-zero; in the finite key setting, of course, it would greatly affect efficiency however we leave that for future work).  Only remaining, now, is to recompute $H(A|B)$.  Let $\tilde{p}_{i,j}$ be the probability that $A$ and $B$ agree on a raw-key bit of $i$ and $j$ respectively in this new lossy scenario.  In this symmetric case, these values are easily found to be:
\begin{align}
\tilde{p}_{0,0} &= \tilde{p}_{1,1} = \frac{1}{M}\left( (1-Q)^2(1-p_l) + p_l(1-Q)\right)\\
\tilde{p}_{0,1} &= \tilde{p}_{1,0} = \frac{1}{M}\left( Q^2(1-p_l) + p_lQ \right),
\end{align}
and where $M$ is the obvious normalization term.  Note that when $p_l = 0$, all equations collapse to the loss-less case as expected.  As we are only interested in the tolerance to noise and loss (i.e., when $r \ge 0$), we do not consider the effective key-rate (as this will only scale the $r$ value in the asymptotic scenario; in the finite key setting, this would not be the case as vacuum events will affect the efficiency of statistic gathering).

\begin{figure}
  \centering
  \includegraphics[width=225pt]{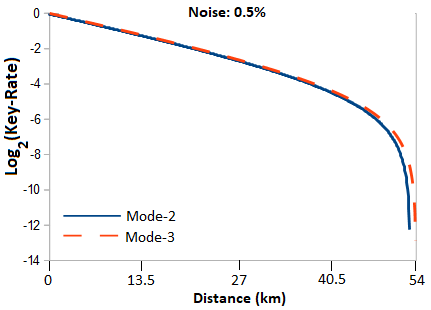}
  \caption{Showing the key-rate of our protocol as a function of distance (in km) for a fixed noise level of $Q = Q_X = .005$.  Here we are using $p_l = 1-10^{-.25 d/10}$, where $d$ is the distance of the channel in km, and $p_l$ is the probability of photon loss in a single channel as described in the text.  Horizontal axis is distance (in km) while vertical axis is log of the key-rate (the ratio of secret key bits to raw-key size).}\label{fig:loss1}
\end{figure}

\begin{figure}
  \centering
  \includegraphics[width=325pt]{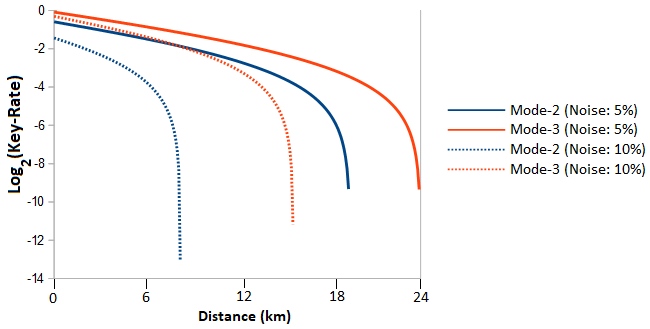}
  \caption{Showing the key-rate of our protocol as a function of distance (in km) for noise levels $Q = Q_X = .05$ and $Q = Q_X = .1$.  Horizontal axis is distance (in km) while vertical axis is log of the key-rate (the ratio of secret key bits to raw-key size).}\label{fig:loss2}
\end{figure}

If we assume a fiber channel, where $1-p_l = 10^{-\alpha \cdot d / 10}$ \cite{QKD-survey}, we may compute the maximal distance over which a key may be distilled assuming this attack model.  These results, for $\alpha = .25dB/km$ are shown in Figures \ref{fig:loss1} and \ref{fig:loss2}.  We observe that, for low-noise levels (Figure \ref{fig:loss1}), the difference between $\MTWO$ and $\MTHREE$ is slight.  However, for higher noise-levels (shown in Figure \ref{fig:loss2}), the difference between the two modes is more substantial.


\section{CLOSING REMARKS}

In this paper, we showed how a semi-quantum protocol may be constructed which, by taking advantage of the two-way quantum channel, can tolerate high levels of noise.  We performed a security analysis and key-rate computation for our protocol comparing to several other (S)QKD protocols.  Our proof of security developed new techniques to take advantage of three-bases with mismatched measurements over two-way channels; these techniques may be applicable to the proof of security of other (S)QKD protocols utilizing two-way quantum communication channels.  We also compared our protocol to BB84 with CAD and showed some interesting similarities and differences in the two settings.

Many interesting future problems remain open.  First, we did not consider finite key settings, and so it would be interesting to consider this.  We utilized numerous mismatched statistics to get our high noise tolerance in the asymptotic setting; while our work here compares favorably with other protocols in the asymptotic setting, it would be interesting to see if this continues to hold true if finite resources are used.  Second, we considered ideal settings; practical devices are only just beginning to be realized in the semi-quantum setting \cite{boyer2017experimentally,krawec2018practical} and it would be interesting to try and adapt some of those techniques to our protocol presented here or, alternatively, to adapt the techniques developed here, to those potentially practical SQKD protocols. Note that, even though these may be practical to implement, standard fully-quantum protocols are still easier to implement thus, this practical study is still mostly of theoretical interest.  Finally, it would be interesting to apply the three-basis mismatched measurement technique we used here to other semi-quantum (or fully-quantum) protocols using a two-way quantum channel; we suspect that improvements to noise tolerances may be established in these cases using the techniques we developed in this paper.

$ $\newline
\textbf{Acknowledgments:} The authors would like to thank the anonymous reviewers for their comments which greatly improved the quality of this paper.
WK would like to acknowledge support from the National Science Foundation under Grant Number 1812070.


\begin{thebibliography}{10}

\bibitem{SQKD-first}
Michel Boyer, Dan Kenigsberg, and Tal Mor.
\newblock Quantum key distribution with classical bob.
\newblock {\em Phys. Rev. Lett.}, 99:140501, Oct 2007.

\bibitem{boyer2017experimentally}
Michel Boyer, Matty Katz, Rotem Liss, and Tal Mor.
\newblock Experimentally feasible protocol for semiquantum key distribution.
\newblock {\em Physical Review A}, 96(6):062335, 2017.

\bibitem{krawec2015security}
Walter~O Krawec.
\newblock Security proof of a semi-quantum key distribution protocol.
\newblock In {\em Information Theory (ISIT), 2015 IEEE International Symposium
  on}, pages 686--690. IEEE, 2015.

\bibitem{krawec2016quantum}
Walter~O Krawec.
\newblock Quantum key distribution with mismatched measurements over arbitrary
  channels.
\newblock {\em Quantum Information and Computation}, 17(3 and 4):209--241,
  2017.

\bibitem{zhang2018security}
Wei Zhang, Daowen Qiu, and Paulo Mateus.
\newblock Security of a single-state semi-quantum key distribution protocol.
\newblock {\em Quantum Information Processing}, 17:1--21, 2018.

\bibitem{QKD-BB84}
Charles~H Bennett and Gilles Brassard.
\newblock Quantum cryptography: Public key distribution and coin tossing.
\newblock In {\em Proceedings of IEEE International Conference on Computers,
  Systems and Signal Processing}, volume 175. New York, 1984.

\bibitem{QKD-LM05}
Marco Lucamarini and Stefano Mancini.
\newblock Secure deterministic communication without entanglement.
\newblock {\em Physical review letters}, 94(14):140501, 2005.

\bibitem{QKD-PingPong}
Kim Bostr{\"o}m and Timo Felbinger.
\newblock Deterministic secure direct communication using entanglement.
\newblock {\em Physical Review Letters}, 89(18):187902, 2002.

\bibitem{QKD-TwoWaySecure}
Normand~J Beaudry, Marco Lucamarini, Stefano Mancini, and Renato Renner.
\newblock Security of two-way quantum key distribution.
\newblock {\em Physical Review A}, 88(6):062302, 2013.

\bibitem{CAD}
Ueli~M Maurer.
\newblock Secret key agreement by public discussion from common information.
\newblock {\em IEEE transactions on information theory}, 39(3):733--742, 1993.

\bibitem{BB84-CAD}
Joonwoo Bae and Antonio Ac{\'\i}n.
\newblock Key distillation from quantum channels using two-way communication
  protocols.
\newblock {\em Physical Review A}, 75(1):012334, 2007.

\bibitem{chau2002practical}
Hoi~Fung Chau.
\newblock Practical scheme to share a secret key through a quantum channel with
  a 27.6\% bit error rate.
\newblock {\em Physical Review A}, 66(6):060302, 2002.

\bibitem{QKD-Tom-1}
Stephen~M Barnett, Bruno Huttner, and Simon~JD Phoenix.
\newblock Eavesdropping strategies and rejected-data protocols in quantum
  cryptography.
\newblock {\em Journal of Modern Optics}, 40(12):2501--2513, 1993.

\bibitem{QKD-Tom-2}
Shun Watanabe, Ryutaroh Matsumoto, and Tomohiko Uyematsu.
\newblock Tomography increases key rates of quantum-key-distribution protocols.
\newblock {\em Physical Review A}, 78(4):042316, 2008.

\bibitem{SQKD-second}
Michel Boyer, Ran Gelles, Dan Kenigsberg, and Tal Mor.
\newblock Semiquantum key distribution.
\newblock {\em Phys. Rev. A}, 79:032341, Mar 2009.

\bibitem{SQKD-lessthan4}
Xiangfu Zou, Daowen Qiu, Lvzhou Li, Lihua Wu, and Lvjun Li.
\newblock Semiquantum-key distribution using less than four quantum states.
\newblock {\em Phys. Rev. A}, 79:052312, May 2009.

\bibitem{SQKD-Single-Security}
Walter~O Krawec.
\newblock Restricted attacks on semi-quantum key distribution protocols.
\newblock {\em Quantum Information Processing}, 13(11):2417--2436, 2014.

\bibitem{krawec2017limited}
Walter~O Krawec and Eric~P Geiss.
\newblock Limited resource semi-quantum key distribution.
\newblock {\em To appear: Proc. ISITA 2018. arXiv preprint arXiv:1710.05076},
  2018.

\bibitem{zou2015semiquantum}
Xiangfu Zou, Daowen Qiu, Shengyu Zhang, and Paulo Mateus.
\newblock Semiquantum key distribution without invoking the classical partyâs
  measurement capability.
\newblock {\em Quantum Information Processing}, 14(8):2981--2996, 2015.

\bibitem{SQKD-MultiUser}
Walter~O Krawec.
\newblock Mediated semiquantum key distribution.
\newblock {\em Physical Review A}, 91(3):032323, 2015.

\bibitem{liu2018mediated}
Zhi-Rou Liu and Tzonelih Hwang.
\newblock Mediated semi-quantum key distribution without invoking quantum
  measurement.
\newblock {\em Annalen der Physik}, 530(4):1700206, 2018.

\bibitem{krawec2018key}
Walter~O Krawec.
\newblock Key-rate bound of a semi-quantum protocol using an entropic
  uncertainty relation.
\newblock In {\em 2018 IEEE International Symposium on Information Theory
  (ISIT)}, pages 2669--2673. IEEE, 2018.

\bibitem{SQKD-secret1}
Qin Li, W.~H. Chan, and Dong-Yang Long.
\newblock Semiquantum secret sharing using entangled states.
\newblock {\em Phys. Rev. A}, 82:022303, Aug 2010.

\bibitem{SQKD-secret2}
Lvzhou Li, Daowen Qiu, and Paulo Mateus.
\newblock Quantum secret sharing with classical bobs.
\newblock {\em Journal of Physics A: Mathematical and Theoretical},
  46(4):045304, 2013.

\bibitem{SQKD-secret3}
Jian Wang, Sheng Zhang, Quan Zhang, and Chao-Jing Tang.
\newblock Semiquantum secret sharing using two-particle entangled state.
\newblock {\em International Journal of Quantum Information}, 10(05), 2012.

\bibitem{SQKD-secret-efficient}
Chun-Wei Yang and Tzonelih Hwang.
\newblock Efficient key construction on semi-quantum secret sharing protocols.
\newblock {\em International Journal of Quantum Information}, 11(05), 2013.

\bibitem{zou2014three}
XiangFu Zou and DaoWen Qiu.
\newblock Three-step semiquantum secure direct communication protocol.
\newblock {\em Science China Physics, Mechanics \& Astronomy},
  57(9):1696--1702, 2014.

\bibitem{shukla2017semi}
Chitra Shukla, Kishore Thapliyal, and Anirban Pathak.
\newblock Semi-quantum communication: protocols for key agreement, controlled
  secure direct communication and dialogue.
\newblock {\em Quantum Information Processing}, 16(12):295, 2017.

\bibitem{yan2018semi}
LiLi Yan, YuHua Sun, Yan Chang, ShiBin Zhang, GuoGen Wan, and ZhiWei Sheng.
\newblock Semi-quantum protocol for deterministic secure quantum communication
  using bell states.
\newblock {\em Quantum Information Processing}, 17(11):315, 2018.

\bibitem{SQKD-comp1}
Kishore Thapliyal, Rishi~Dutt Sharma, and Anirban Pathak.
\newblock Orthogonal-state-based and semi-quantum protocols for quantum private
  comparison in noisy environment.
\newblock {\em arXiv preprint arXiv:1608.00101}, 2016.

\bibitem{SQKD-comp2}
Wen-Han Chou, Tzonelih Hwang, and Jun Gu.
\newblock Semi-quantum private comparison protocol under an almost-dishonest
  third party.
\newblock {\em arXiv preprint arXiv:1607.07961}, 2016.

\bibitem{SQKD-comp3}
Lang Yan-Feng.
\newblock Semi-quantum private comparison using single photons.
\newblock {\em International Journal of Theoretical Physics},
  57(10):3048--3055, 2018.

\bibitem{SQKD-comp4}
Tian-Yu Ye and Chong-Qiang Ye.
\newblock Measure-resend semi-quantum private comparison without entanglement.
\newblock {\em International Journal of Theoretical Physics},
  57(12):3819--3834, 2018.

\bibitem{krawec2018practical}
Walter~O Krawec.
\newblock Practical security of semi-quantum key distribution.
\newblock In {\em Quantum Information Science, Sensing, and Computation X},
  volume 10660, page 1066009. International Society for Optics and Photonics,
  2018.

\bibitem{SQKD-prac3}
Michel Boyer, Rotem Liss, and Tal Mor.
\newblock Attacks against a simplified experimentally feasible semiquantum key
  distribution protocol.
\newblock {\em Entropy}, 20(7), 2018.

\bibitem{SQKD-photon-tag}
Yong-gang Tan, Hua Lu, and Qing-yu Cai.
\newblock Comment on Òquantum key distribution with classical bobÓ.
\newblock {\em Phys. Rev. Lett.}, 102:098901, Mar 2009.

\bibitem{SQKD-photon-tag-comment}
Michel Boyer, Dan Kenigsberg, and Tal Mor.
\newblock Boyer, kenigsberg, and mor reply:.
\newblock {\em Phys. Rev. Lett.}, 102:098902, Mar 2009.

\bibitem{CV2}
Carlo Ottaviani and Stefano Pirandola.
\newblock General immunity and superadditivity of two-way gaussian quantum
  cryptography.
\newblock {\em Scientific reports}, 6:22225, 2016.

\bibitem{CV4}
Carlo Ottaviani, Stefano Mancini, and Stefano Pirandola.
\newblock Two-way gaussian quantum cryptography against coherent attacks in
  direct reconciliation.
\newblock {\em Physical Review A}, 92(6):062323, 2015.

\bibitem{CV5}
Quntao Zhuang, Zheshen Zhang, Norbert L{\"u}tkenhaus, and Jeffrey~H Shapiro.
\newblock Security-proof framework for two-way gaussian
  quantum-key-distribution protocols.
\newblock {\em Physical Review A}, 98(3):032332, 2018.

\bibitem{CV6}
Shouvik Ghorai, Eleni Diamanti, and Anthony Leverrier.
\newblock Composable security of two-way continuous-variable quantum key
  distribution without active symmetrization.
\newblock {\em Physical Review A}, 99(1):012311, 2019.

\bibitem{CV-floodlight}
Quntao Zhuang, Zheshen Zhang, Justin Dove, Franco N.~C. Wong, and Jeffrey~H.
  Shapiro.
\newblock Floodlight quantum key distribution: A practical route to
  gigabit-per-second secret-key rates.
\newblock {\em Phys. Rev. A}, 94:012322, Jul 2016.

\bibitem{CV1}
Stefano Pirandola, Stefano Mancini, Seth Lloyd, and Samuel~L Braunstein.
\newblock Continuous-variable quantum cryptography using two-way quantum
  communication.
\newblock {\em Nature Physics}, 4(9):726, 2008.

\bibitem{CV3}
Christian Weedbrook, Carlo Ottaviani, and Stefano Pirandola.
\newblock Two-way quantum cryptography at different wavelengths.
\newblock {\em Physical Review A}, 89(1):012309, 2014.

\bibitem{QKD-survey}
Valerio Scarani, Helle Bechmann-Pasquinucci, Nicolas~J. Cerf, Miloslav
  Du\ifmmode~\check{s}\else \v{s}\fi{}ek, Norbert L\"utkenhaus, and Momtchil
  Peev.
\newblock The security of practical quantum key distribution.
\newblock {\em Rev. Mod. Phys.}, 81:1301--1350, Sep 2009.

\bibitem{QKD-symmetric}
Robert Konig and Renato Renner.
\newblock A de finetti representation for finite symmetric quantum states.
\newblock {\em Journal of Mathematical physics}, 46:122108, 2005.

\bibitem{QKD-general-attack}
Matthias Christandl, Robert Konig, and Renato Renner.
\newblock Postselection technique for quantum channels with applications to
  quantum cryptography.
\newblock {\em Phys. Rev. Lett.}, 102:020504, Jan 2009.

\bibitem{QKD-general-attack2}
Renato Renner.
\newblock Symmetry of large physical systems implies independence of
  subsystems.
\newblock {\em Nature Physics}, 3(9):645--649, 2007.

\bibitem{QKD-Winter-Keyrate}
Igor Devetak and Andreas Winter.
\newblock Distillation of secret key and entanglement from quantum states.
\newblock {\em Proceedings of the Royal Society A: Mathematical, Physical and
  Engineering Science}, 461(2053):207--235, 2005.

\bibitem{QKD-renner-keyrate}
Renato Renner, Nicolas Gisin, and Barbara Kraus.
\newblock Information-theoretic security proof for quantum-key-distribution
  protocols.
\newblock {\em Phys. Rev. A}, 72:012332, Jul 2005.

\bibitem{QKD-BB84-Modification}
Hoi-Kwong Lo, Hoi-Fung Chau, and M~Ardehali.
\newblock Efficient quantum key distribution scheme and a proof of its
  unconditional security.
\newblock {\em Journal of Cryptology}, 18(2):133--165, 2005.

\bibitem{Krawec2016Reflect}
Walter~O. Krawec.
\newblock Security of a semi-quantum protocol where reflections contribute to
  the secret key.
\newblock {\em Quantum Information Processing}, 15(5):2067--2090, May 2016.

\bibitem{lucamarini2014quantum}
Marco Lucamarini and Stefano Mancini.
\newblock Quantum key distribution using a two-way quantum channel.
\newblock {\em Theoretical Computer Science}, 560:46--61, 2014.

\end{thebibliography}

\end{document}